\documentclass[aps,nofootinbib,letterpaper,onecolumn,notitlepage,12pt,raggedbottom]{revtex4-1}
\usepackage{graphicx,amssymb,amsmath,fullpage}
\usepackage{slashed}
\usepackage{setspace}
\usepackage{enumerate}
\usepackage{color}
\usepackage[usenames,dvipsnames]{xcolor}
\usepackage{hyperref}
\usepackage{bm}
\usepackage{epstopdf}
\usepackage{feynmp}
\usepackage{graphicx}
\usepackage{epstopdf}

 \makeatletter
\def\l@subsubsection#1#2{}
\makeatother

\usepackage{tikz}
\usetikzlibrary{arrows} 
\usetikzlibrary{decorations.pathmorphing} 
\usetikzlibrary{decorations.markings} 

\tikzset{
vector/.style={decorate, decoration={snake}, draw},
doubvector/.style={decorate, decoration={snake}, draw,double distance=1.4pt},
fermion/.style={draw=black},
fwfermion/.style={draw=black, postaction={decorate},decoration={markings,mark=at position .55 with {\arrow[very thick]{latex'}}}},
bwfermion/.style={draw=black, postaction={decorate},decoration={markings,mark=at position .55 with {\arrow[very thick]{latex' reversed}}}},
doubfermion/.style={draw=black,double distance=1.4pt},
fwdoubfermion/.style={draw=black,double distance=1.4pt,postaction={decorate},decoration={markings,mark=at position .55 with {\arrow[very thick]{latex'}}}},
bwdoubfermion/.style={draw=black,double distance=1.4pt,postaction={decorate},decoration={markings,mark=at position .55 with {\arrow[very thick]{latex' reversed}}}},
gluon/.style={decorate, draw=black,decoration={coil,amplitude=3pt,segment length=4pt}},
scalar/.style={dashed,draw=black},
fwscalar/.style={dashed,draw=black,postaction={decorate},decoration={markings,mark=at position .55 with {\arrow[very thick]{latex'}}}},
bwscalar/.style={dashed,draw=black,postaction={decorate},decoration={markings,mark=at position .55 with {\arrow[very thick]{latex' reversed}}}},
doubscalar/.style={dashed,draw=black,double distance=1.4pt},
fwdoubscalar/.style={dashed,draw=black,double distance=1.4pt,postaction={decorate},decoration={markings,mark=at position .55 with {\arrow[very thick]{latex'}}}},
bwdoubscalar/.style={dashed,draw=black,double distance=1.4pt,postaction={decorate},decoration={markings,mark=at position .55 with {\arrow[very thick]{latex' reversed}}}},
}

\numberwithin{equation}{section}

\def\app#1#2{%
  \mathrel{%
    \setbox0=\hbox{$#1\sim$}%
    \setbox2=\hbox{%
      \rlap{\hbox{$#1\propto$}}%
      \lower1.1\ht0\box0%
    }%
    \raise0.25\ht2\box2%
  }%
}

\newcommand{\Sec}[1]{Sec. \ref{#1}}
\newcommand{\Fig}[1]{Fig. \ref{#1}}
\newcommand{\Tab}[1]{Table \ref{#1}}
\newcommand{\Equ}[1]{Eq. (\ref{#1})}

\def\gsim{\mathrel{\rlap{\lower4pt\hbox{$\sim$}}
    \raise1pt\hbox{$>$}}}

\newcommand*\xbar[1]{%
  \hbox{%
    \vbox{%
      \hrule height 0.4pt 
      \kern0.3ex
      \hbox{%
        \kern-0.05em
        \ensuremath{#1}%
        \kern-0.05em
      }%
    }%
  }%
} 

\newcommand*{\antiBox}{\xbar{\Box}}

\begin{document}

\title{A 2 TeV $W_R$, Supersymmetry, and the Higgs Mass}
\author{Jack H. Collins}
\email{jhc296@cornell.edu}
\affiliation{Department of Physics, LEPP, Cornell University, Ithaca, NY 14853, USA}
\author{Wee Hao Ng}
\email{wn68@cornell.edu}
  \affiliation{Department of Physics, LEPP, Cornell University, Ithaca, NY 14853, USA}
\date{\today}

\begin{abstract}
\begin{spacing}{1.}
A recent ATLAS search for diboson resonances and a CMS search for $eejj$ resonances which both show excesses with significance around $3 \sigma$ have generated interest in $SU(2)_R$ gauge extensions of the Standard Model with a $W'$ mass around 2 TeV. We investigate the possibility that an $SU(2)_R$ gauge extension of the MSSM compatible with an explanation of the diboson anomaly might give rise to a significant enhancement of the Higgs mass above the MSSM tree level bound $m_{h, \text{tree}} < 90 \; \text{GeV}$ due to non-decoupling $D$-terms. This model contains a vector-like charge $-1/3$ $SU(2)_R$ singlet quark for each generation which mixes significantly with the $SU(2)_R$ doublet quarks, affecting the $W_R$ phenomenology. We find that it is possible to achieve $m_{h, \text{tree}} > 110 \; \text {GeV}$, and this requires that the $Z'$ mass is close to 3 TeV.
\end{spacing}
\end{abstract}

\maketitle


\section{Introduction}
The recently discovered Higgs boson with mass around 125~GeV creates some tension in the Minimal Supersymmetric Standard Model (MSSM). This is because its quartic interaction comes only from its supersymmetric gauge interactions at tree level, resulting in the well known result that at tree level the Higgs mass is no greater than the $Z$ boson mass of 91~GeV.
\begin{equation}
m_{h, \text{tree}}^2 = \frac{1}{4}(g^2 + g'^2)v^2 \cos^2\left(2 \beta\right) \leq m_Z^2
\end{equation}
Evading this constraint with minimal matter content requires significant radiative corrections from stop loops, necessitating some combination of a large soft SUSY breaking mass and large $A$-terms. This in turn incurs a large fine tuning penalty in the Higgs potential due to the quadratic sensitivity of the Higgs soft mass to these parameters. It is possible that this little hierarchy problem is resolved by extending the matter content of the MSSM to allow for new tree level contributions to the Higgs quartic, either from $F$-terms as in the NMSSM \cite{Ellis:1988er,Drees:1988fc}, indicating the presence of new chiral superfields, or from new $D$-term contributions as is possible in gauge extensions of the MSSM \cite{Batra:2003nj,Maloney:2004rc,Bertuzzo:2014sma}. The latter possibility predicts the existence of heavy gauge boson resonances that may be observable at the LHC.

With this in mind, it is intriguing that a number of small anomalies with local significance of up to $3.4 \sigma$ have been reported by the ATLAS and CMS experiments which might speculatively be interpreted as resulting from a new resonance with mass 1.8 -- 2~TeV. The most significant excess is in an ATLAS search for resonances decaying in pairs of SM vector bosons (either $W$ or $Z$) which in turn decay hadronically \cite{Aad:2015owa}, finding a maximum local significance of $3.4 \sigma$ and limits weaker than expected for diboson resonances with masses between 1.8 and 2.2~TeV\footnote{See also \cite{Allanach:2015hba,Fichet:2015yia} for a detailed discussion of this excess.}. However, their leptonic and semileptonic searches for diboson resonances which have a similar sensitivity in this mass range saw no deviation from SM expectations \cite{Aad:2015ufa, Aad:2014xka}. A combination of these ATLAS searches finds a maximum significance of $2.5\sigma$, with limits weaker than expected in the mass window 1.9 -- 2.1~TeV \cite{ATLASdibosoncomb}. A CMS search for hadronically decaying diboson resonances saw a much smaller excess of 1 -- 1.5$\sigma$ in the mass window 1.8 -- 2.0 TeV \cite{Khachatryan:2014hpa}, and their semileptonic search for a leptonically decaying $Z$ and a hadronically decaying vector boson found a $1.5 \sigma$ excess in the mass window 1.7 -- 1.9~TeV \cite{Khachatryan:2014gha}. A CMS search for $WH$ resonances decaying into $l \nu b b$ found a $1.9\sigma$ excess in the mass window 1.8 -- 2~TeV. In addition, CMS and ATLAS find modest excesses in their dijet mass distributions in the window 1.7 -- 1.9~TeV with significance $2.2\sigma$ and $1 \sigma$ respectively \cite{Khachatryan:2015sja, Aad:2014aqa}. Finally, a CMS search in the $e e j j $ final state found a $2.8 \sigma$ excess consistent with being produced by a resonance in the mass range 1.8 -- 2.2 TeV \cite{Khachatryan:2014dka}.

It has been pointed out that a compelling candidate to explain these anomalies, if they are indeed a first hint of new physics, is a $W'$ from a broken gauge symmetry which couples to right handed (RH) currents \cite{Brehmer:2015cia, Hisano:2015gna, Cheung:2015nha, Dobrescu:2015qna, Gao:2015irw, Cao:2015lia, Dobrescu:2015yba, Krauss:2015nba, Dev:2015pga, Coloma:2015una, Gluza:2015goa,Deppisch:2015cua}, as in models with Left-Right symmetry (LRS) \cite{Pati:1974yy, Mohapatra:1974hk}. Firstly, such a particle is not constrained by the strong limits on $l^+ l^-$ or $l \nu$ resonances if it is charged and does not have a significant coupling to LH leptons. Secondly, the $eejj$ excess might be explained by a decay chain via RH neutrinos, $W_R \to e_R \nu_R \to e_R e_R j j$ \cite{Keung:1983uu}. The possibility of a new gauge symmetry is exciting in and of itself, but it could have a very special significance in the context of a supersymmetric theory due to the interplay between gauge symmetries and the Higgs mass. The purpose of this paper is to explore the possibility that these anomalies could be directly related to the Higgs mass. We therefore consider a model with 1.9 TeV $W_R$ with properties necessary to explain the anomalies.

The simplest possibility for electroweak symmetry breaking (EWSB) in these models is that it is generated by the vevs of a bidoublet under $SU(2)_L \times SU(2)_R$, which contains the $H_u$, $H_d$ fields of the MSSM with vevs $v_u$, $v_d$. This provides the $W_L\text{--} W_R$ mixing that is necessary for the diboson decay signature. As we shall review in more detail in \Sec{sec:model}, the $D$-term contribution to the Higgs mass in these models is given by \cite{Zhang:2008jm, Babu:2014vba}
\begin{equation}
m_{h, \text{tree}}^2 \leq \frac{1}{4}\left(g^2 + g_R^2 \right) v^2 \cos^2\left(2 \beta \right),
\end{equation} 
where $\tan \beta = v_u / v_d$ as in the MSSM. Large contributions to the Higgs mass therefore require large $g_R$ and large $\tan \beta$. In a minimal model this is not possible to reconcile with the anomalies. This is because the partial width $\Gamma(W' \to W Z)$ is suppressed by a factor $\sin^2 (2 \beta) / 24$ compared to the partial width into dijets. A recent paper \cite{Brehmer:2015cia} fitted the cross sections for the dijet and diboson signatures, and found that 
\begin{equation}
\frac{\sigma \times \text{BR}(W' \to WZ)}{\sigma \times \text{BR}(W' \to jj)} = \frac{\sin^2 2 \beta}{24} > \frac{2.4\;\text{fb}}{144 \; \text{fb}}.
\label{equ:WZoverjj}
\end{equation}
Satisfying this inequality requires $\tan \beta \simeq 1$. Furthermore, fitting the overall signal cross section requires $g_R / g < 0.8$ in minimal models \cite{Brehmer:2015cia, Hisano:2015gna, Cheung:2015nha, Dobrescu:2015qna, Gao:2015irw, Cao:2015lia, Dobrescu:2015yba, Krauss:2015nba, Dev:2015pga, Coloma:2015una,Gluza:2015goa, Deppisch:2015cua}, since $\sigma_{W'} \propto g_R^2$. Fitting the excess with larger $\tan \beta$ and $g_R$ therefore requires a departure from minimality. This might be possible by suppressing the $W_R$ coupling to the RH quark doublets, which would modify the Drell-Yan production cross section and the inequality of \Equ{equ:WZoverjj}. In this paper we achieve this by introducing a vector-like charge $-1/3$ quark for each generation which mixes with the $SU(2)_R$ quark doublets after that gauge symmetry is broken. The right handed down-type quarks of the SM are then admixtures from the $SU(2)_R$ doublets and the singlets, with some mixing angle $\theta_d$. The $W_R u_R d_R$ coupling is then suppressed by a factor of $\cos \theta_d$. Varying this mixing angle allows the freedom to fit the data with a larger value of $\tan \beta$, and since $\sigma_{W'} \propto g_R^2 \cos^2 \theta_d$, a smaller $\cos \theta_d$ also allows the excess to be fit with a larger $g_R$. It is worth noting that while we introduce these new fields for purely phenomenological purposes, they are expected in E6 GUTs \cite{Hewett:1988xc}. We do not explore the neutrino sector in this paper, and therefore do not discuss the $eejj$ signature in any detail. The collider phenomenology of the right handed neutrinos might be modified by light electroweak SUSY states such as Higgsinos as has been discussed in some detail in a recent paper \cite{Krauss:2015nba}.

We describe the model in \Sec{sec:model}, where we also review non-decoupling D-terms and the relevant experimental data. The main results of our paper -- the implications for the Higgs mass in our model -- are presented in \Sec{sec:results}. The couplings associated with the new quark fields are strongly constrained by flavour changing neutral current (FCNC) observables, which we discuss in \Sec{sec:flavour}. Finally, we review the main conclusions of this work in \Sec{sec:conclusions}.

\section{\label{sec:model}The Model}

We work with the gauge group $SU(3)_c \times SU(2)_L \times SU(2)_R \times U(1)_X$, with a symmetry breaking $SU(2)_R \times U(1)_X \to U(1)_Y$ at $\sim 2 \; \text{TeV}$. The chiral superfields of the model are summarized in \Tab{tab:fields}. In general, the RH gauge symmetry might be broken by some combination of doublet and triplet vevs
\begin{equation}
H_R = \begin{pmatrix}H_R^+ \\ \frac{v_D}{\sqrt{2}} + H_R^0\end{pmatrix}, ~~~ \Delta = \begin{pmatrix}\frac{1}{2}\Delta^+&\Delta^{++}\\\frac{v_\Delta}{\sqrt{2}} + \Delta^0& -\frac{1}{2}\Delta^+ \end{pmatrix},  ~~~ 
\xbar{\Delta} = \begin{pmatrix}\frac{1}{2}\xbar{\Delta}^-& \frac{v_{\bar{\Delta}}}{\sqrt{2}} + \xbar{\Delta}^0\\ \xbar{\Delta}^{--}& -\frac{1}{2}\xbar{\Delta}^- \end{pmatrix}.\label{equ:vevs}
\end{equation}
The $H_R$ might be identified with a RH lepton doublet, or else must come with a conjugate superfield with opposite $X$ charge for anomaly cancellation. For simplicity we assume such a field does not acquire a significant vev, though this would not significantly alter our conclusions. The unbroken hypercharge generator is given by
\begin{equation}
Y = T^3_R + X, ~~~~~g'^{-2} = g_R^{-2} + g_X^{-2}.
\end{equation}
Writing $v_T^2 = v_\Delta^2 + v_{\bar{\Delta}}^2$, the $W'$ and $Z'$ masses are given by
\begin{align}
m_{W'}^2 &= \frac{1}{4}g_R^2 \left(2 v_T^2 + v_D^2 \right)\left(1 + \mathcal{O}\left(\frac{v^2}{2 v_T^2 + v_D^2} \right) \right)\\
m_{Z'}^2 &= \frac{1}{4}\left(g_R^2 + g_X^2 \right)\left(4 v_T^2 + v_D^2 \right)\left(1 + \mathcal{O}\left(\frac{v^2}{4 v_T^2 + v_D^2} \right) \right)
\end{align}
with $v = 246 \; \text{GeV}$ the EWSB vev. By analogy with EWSB, the relation between the $W'$ and $Z'$ masses can be parametrized in terms of a new Weinberg angle, $\theta_{w'}$, and $\rho'$ parameter
\begin{equation}
\frac{m_{Z'}^2}{m_{W'}^2} = \frac{\rho'}{c_{w'}^2} \label{equ:rhoprime}
\end{equation}
with
\begin{equation}
1 \leq \rho' \leq 2 ~~~~~~~~~~
 c_{w'}^2 \equiv \frac{g_R^2}{g_R^2 + g_X^2} = 1 - \frac{s_w^2 g^2}{c_w^2 g_R^2}.
\end{equation}
For pure doublet breaking $\rho' = 1$ as in the SM, while for pure triplet breaking $\rho' = 2$.

\begin{table}
\center
\begin{tabular}{l|cccc}
&$SU(3)_c$&$SU(2)_L$&$SU(2)_R$&$U(1)_X$\\ \hline
$Q_{L i} = ({u_L}_{i}, {d'_L}_{i})$ & $\Box$ & $\Box$ & $\mathbf{1}$ & $1/6$\\
$Q^c_{R i} = ({d'^c_R}_{i},{u^c_R}_{i})$ & $\antiBox$ & $\mathbf{1}$ & $\Box$ & $-1/6$\\
${D'}_i$ & $\Box$ & $\mathbf{1}$ & $\mathbf{1}$ & $-1/3$\\
${D'}^c_i$ & $\antiBox$ & $\mathbf{1}$ & $\mathbf{1}$ & $1/3$\\
$L_{L i} = (\nu_{L i}, \ell_{L i})$ & $\mathbf{1}$ & $\Box$ & $\mathbf{1}$ & $-1/2$\\
$L_{R i} = (\ell_{R i},\nu_{R i})$ & $\mathbf{1}$ & $\mathbf{1}$ & $\Box$ & $1/2$\\
$\Phi = (H_u, H_d)$ & $\mathbf{1}$ & $\Box$ & $\Box$ & $0$\\
$\Delta , \xbar{\Delta}$ & $\mathbf{1}$ & $\mathbf{1}$ & $\mathbf{3}$ & $\pm 1$\\
$H_{R}$ & $\mathbf{1}$ & $\mathbf{1}$ & $\Box$ & $1/2$
\end{tabular}
\caption{Chiral superfields.}
\label{tab:fields}
\end{table}

If EWSB is achieved with by a bidoublet $\Phi = (H_u, H_d)$ with vevs $v_u/\sqrt{2}, v_d/\sqrt{2}$ and $v^2 = v_u^2 + v_d^2$ then the  $W_L\text{--}W_R$ mass matrix is given by
\begin{equation}
M_{W, LR}^2 = \frac{1}{4}\begin{pmatrix}g^2 v^2 & -2g g_R v_u v_d^*\\
 - 2g g_R v_u^* v_d& g_R^2 \left(2 v_T^2 + v_D^2 + v^2\right)\end{pmatrix}.
\end{equation}
This matrix is diagonalised with a rotation angle
\begin{equation}
\sin \phi \simeq \frac{g_R}{g} \frac{m_W^2}{m_{W'}^2} \sin 2 \beta,
\end{equation}
with $\tan \beta = v_u / v_d$. The decay responsible for the diboson signature, $W' \to W Z$, has a width given by
\begin{align}
\Gamma(W' \to W Z) &\simeq \frac{m_{W'}^5}{192 \pi m_W^2 m_Z^2} \frac{g^2}{c_w^2} \sin^2 \phi \notag\\
&= \frac{m_{W'}}{192 \pi} g_R^2 \sin^2 2 \beta,\label{equ:dibosonwidth}
\end{align}
which can be calculated from the kinetic terms of the Lagrangian \cite{Brehmer:2015cia, Deshpande:1988qq}. The diboson signature is therefore maximised for $v_u \simeq v_d$ and hence $\sin 2 \beta \simeq 1$.

\subsection{Non-Decoupling D-terms}
In this model, the $D$-terms in the Higgs sector are given by
\begin{align}
V_D = &\frac{g^2}{8}\left| \Phi^\dagger \sigma_a \Phi \right|^2 + \frac{g_R^2}{8}\left|H_R^\dagger \sigma_a H_R + 2\Delta^\dagger \sigma_a \Delta + 2 \xbar{\Delta}^\dagger \sigma_a \xbar{\Delta} + \Phi \sigma_a \Phi^\dagger \right|^2 \notag\\
&+ \frac{g_X^2}{8} \left|2\Delta^\dagger \Delta - 2\xbar{\Delta}^\dagger \xbar{\Delta} + H_R^\dagger H_R \right|^2
\end{align}
Substituting in the vevs of \Equ{equ:vevs} and focussing on the terms relevant for the calculation of the potential for the neutral EWSB Higgses, we arrive at
\begin{align}
V_D \supset &\frac{1}{8}\left(g^2 + g_R^2 \right) \left(\left|H_u^0\right|^2 -  \left|H_d^0\right|^2\right)^2 \notag \\
& + \frac{g_R^2}{2}\text{Re}\left(2 v_\Delta \Delta^0 - 2 v_{\bar{\Delta}} \xbar{\Delta}^0 + v_D H_R^0\right)\left(\left|H_u^0\right|^2 -  \left|H_d^0\right|^2\right).
\end{align}
The effective $D$-term for the MSSM-like Higgs fields is obtained by adding the first term from the equation above with the term obtained by integrating out the linear combination $\text{Re} (2 v_\Delta \Delta^0 - 2 v_{\bar{\Delta}} \xbar{\Delta}^0 + v_D H_R^0)$. This field is the scalar superpartner of the Goldstone which is eaten by the $Z'$, and in the supersymmetric limit in which this symmetry breaking occurs far above the scale of supersymmetry breaking the mass of this field is the same as that of the $Z'$ and integrating it out returns the classic MSSM result, $V_D = (g^2 + g'^2)(|H_u^0|^2 - |H_d^0|^2)/8$ \cite{Batra:2003nj, Maloney:2004rc,Bertuzzo:2014sma}. 

In the case that $m_\text{SUSY} \sim m_{Z'}$ that we will be considering in this paper, this scalar will gain an additional SUSY breaking contribution to its mass that is important for calculating the effective quartic for the EWSB Higgses. The general result is that the tree level Higgs mass contribution from $D$-terms is given by
\begin{equation}
m_{h, \text{tree}}^2 = \frac{1}{4}\left(g^2 + \xi g_R^2 \right) v^2 \cos^2 2 \beta, ~~~~~~ \xi = 1-\frac{g_R^2}{g_R^2 + g_X^2 + \delta}.\label{equ:higgsmassbound}
\end{equation}
Any model dependence is encoded in the parameter $\delta$, which interpolates between the decoupling limit ($\delta \to 0$) and the non-decoupling limit ($\delta \to \infty$)\footnote{We have implicitly assumed that the decoupling limit exists in this discussion.}. The relation between $\delta$ and the paramers of the scalar potential is generically of the form $\delta \sim m_0^2 / v_R^2$, where $m_0$ is the typical scale of the SUSY breaking parameters in the $SU(2)_R$ Higgs sector. The precise form of this relationship will be model dependent, but larger values of $\delta$ will generically correspond to a greater degree of tuning in the $SU(2)_R$ breaking potential. We discuss a simple model of triplet breaking in appendix \ref{sec:tuning} which illustrates the main points. For our numerical work in the next section, we take as benchmark points the values $\delta = \infty$ and $\delta = 2.5$ to describe tuned and untuned scenarios respectively.

As in the MSSM, the $D$-term contribution to the Higgs mass is maximised for $\cos 2 \beta = 1$, while the diboson signature is maximised is for $\sin 2 \beta = 1$, \Equ{equ:dibosonwidth}. This is a key tension in trying to reconcile the diboson signature with large non-decoupling $D$-terms.

\subsection{\label{sec:exotic}Exotic Quarks}

The couplings of the quarks to the Higgses are given by the superpotential
\begin{equation}
W \supset y Q_L \Phi Q_R^c + z D' H_R Q_R^c + M D' D'^c \label{equ:exoticsuperpotential}
\end{equation}
where $y$, $z$, $M$ are matrices in flavour-space. After the breaking of $SU(2)_R$ but before EWSB, a linear combination of $d'^c_R, D'^c$ marries the field $D'$ and obtain a large Dirac mass, $m_D \simeq \sqrt{(z v_D)^2 /2 + M^2}$, with the remaining linear combination remaining massless and which can be identified with the RH down-type quarks of the SM, $d_R^c$. We can write
\begin{equation}
\begin{pmatrix}d_R^c\\s_R^c\\b_R^c\end{pmatrix} \simeq \begin{pmatrix}c_d&&\\&c_d&\\&&c_b\end{pmatrix} \begin{pmatrix}d'^c_R\\s'^c_R\\b'^c_R \end{pmatrix}+ \begin{pmatrix}s_d&&\\&s_d&\\&&s_b\end{pmatrix} \begin{pmatrix}{D'}^c\\{S'}^c\\{B'}^c\end{pmatrix}\label{equ:mixing}
\end{equation}
with $c_d = \cos \theta_d, s_d = \sin \theta_d$, and $\tan \theta_d \sim z_{11} v_d/(\sqrt{2} M_{11})$. In the limit $M \to \infty$ we recover the structure of a minimal left-right symmetric model, in which the RH down-type quarks are $SU(2)_R$ partners of the RH up-type quarks and $s_d, s_b \to 0$. In order to evade constraints from FCNCs, we have assumed that the upper left $2\times2$ block of the rotation matrix is close to the identity matrix and the mixing between the third and first two generations are small. This structure might be enforced by an approximate $U(2) \times U(1)$ flavour symmetry. We will explore the constraints on this flavour structure in more detail in \Sec{sec:flavour}.

Because the up and down type quarks couple to the bidoublet with the same Yukawa matrix $y$, the expectation from Eqs. (\ref{equ:exoticsuperpotential}) and (\ref{equ:mixing}) is that their masses have the relationship
\begin{equation}
\frac{m_u}{m_d} \simeq \frac{m_c}{m_s} \simeq \frac{\tan \beta}{c_d}, ~~~~~~ \frac{m_t}{m_b} \simeq \frac{\tan \beta}{c_b}.
\end{equation}
The mass relationships for the light quarks might easily be modified without introducing large FCNCs either as a result of additional loop contributions from the squark sector \cite{Babu:1998tm}, or from additional small sources of EWSB which couple to the first and second generation quarks via non-renormalizable operators \cite{Guadagnoli:2010sd}. However, it is difficult to account for the mass ratio for the third generation quarks with small $\tan \beta$ and $c_b = 1$ by altering the EWSB sector without also suppressing the diboson signature. On the other hand, this mass ratio is well accounted for if $c_b \simeq t_\beta m_b / m_t \simeq t_\beta / 35$. We will assume this relationship in this paper. This means that $b_R^c$ is mostly an $SU(2)_R$ singlet and the partial width for $W' \to t b$ is suppressed by a factor $c_b^2$. On the other hand, due to the potential sensitivity of the light quark masses to other small corrections we do not use these mass ratios to constrain $c_d$.

As a consequence of this mixing, the production cross section of the $W'$ and its partial width into dijets are modified:
\begin{align}
\sigma_{W'} &\propto c_d^2 \, g_R^2\label{equ:suppressedXS},\\
\Gamma\left(W' \to j j\right) &= \frac{m_{W'}}{8 \pi}c_d^2 \, g_R^2,\\
\frac{\Gamma\left(W' \to WZ\right)}{\Gamma\left(W' \to j j\right)} &= \frac{\sin^2 2 \beta}{24 \, c_d^2}.\label{equ:enhancedWZBR}
\end{align}
A smaller $c_d$ allows for a larger diboson branching fraction, providing the freedom to lower $\sin^2 2 \beta$, due to \Equ{equ:enhancedWZBR}. It also allows the same $W'$ cross section to be achieved with a larger $g_R$ due to \Equ{equ:suppressedXS}. The combination of these factors is what allows for an enhancement of the tree level Higgs mass in \Equ{equ:higgsmassbound} compared to the minimal model which corresponds to $c_d = 1$. It is worth bearing in mind that while we are mainly driven by the relation between the experimental excesses and the Higgs mass, the region of parameter space near $c_d \simeq t_\beta \, m_s / m_c \simeq t_\beta / 14$ might be particularly interesting for flavour physics.

It is expected that the first and second generation exotic quarks $D, S$ would decay via $D \to Z j$, $D \to W j$ with significant branching fractions via $W' \text{--} W$ and $Z' \text{--} Z$ mixing. Only one dedicated LHC search exists for this scenario, a search for $Q \to W q$ by the ATLAS experiment \cite{Aad:2015tba}. They found a broad $2 \sigma$ excess, and excluded the mass range $320 \; \text{GeV}$ to $690 \; \text{GeV}$ for $\text{BR}(Q \to W q) = 100\%$. There are no exclusions if this branching ratio is less than $40 \%$. On the other hand, there are a variety of searches by both the ATLAS and CMS collaborations for bottom quark partners decaying via $B \to h b$, $B \to Z b$, $B \to W t$ \cite{Aad:2015kqa, Aad:2015gdg,Aad:2014efa,Aad:2015mba,Khachatryan:2015gza}. The strongest bounds were set by CMS, which found upper limits on the mass of the bottom partner ranging between 750 GeV and 900 GeV depending on its branching ratios. Giving the bottom partner a sufficiently large mass to evade these limits requires $v_D \gtrsim 1 \; \text{TeV}$ if the theory is weakly coupled. Since we are allowing for a dominantly triplet-breaking scenario with $\rho' \simeq 2$ in our analysis, it needs to be checked that this is compatible with a TeV scale doublet vev. Indeed, setting $g_R = g$ and $v_D = 1 \; \text{TeV}$ results in $\rho' = 1.97$ and $v_T = 4.0 \; \text{TeV}$, while for $g_R = 1.4 \, g$ we get $\rho' = 1.94$ and $v_T = 2.8 \; \text{TeV}$. It is therefore compatible to take $\rho \simeq 2$ while assuming the vector-like quarks are heavy and mix significantly with the doublet quarks.

\section{\label{sec:results}Results and Discussion}

In this section we explore the parameter space of the model in order to find regions that can explain 2 TeV anomalies and generate a large $D$-term contribution for the Higgs mass without being excluded by other searches. The main parameters controlling the $W'$ signature in the diboson and dijet channels are $g_R$, $c_d^2$, $\tan \beta$. In this section we choose to set $\text{BR}(W' \to \text{SM}) = 100 \%$ for simplicity of the analysis. Additional decays are possible into $\ell_R \nu_R$ (which might be responsible for the $eejj$ excess), into exotic quarks and into squarks and other SUSY states. We provide a brief discussion of these effects in \Sec{sec:couplings} and \Fig{fig:extraBRplot}. Important constraints on the parameter space will come from limits on the mass and couplings of the $Z'$ due to LHC resonance searches and due to electroweak precision constraints. This makes the parameter $\rho'$ relevant to the analysis. Additionally, the Higgs mass depends on the parameter $\delta$ which we will take as either 2.5 or $\infty$. We use the fits to the $W'$ diboson and dijet signatures provided in \cite{Brehmer:2015cia}. The $W'$ and $Z'$ cross sections and branching ratios are calculated using the couplings listed in Appendix \ref{sec:couplings} and the NNPDF2.3 PDF set \cite{Ball:2012cx}. The parameter ranges considered in this analysis are summarized in \Tab{tab:params}.

\begin{table}
\center
\begin{tabular}{c l c l}
\textbf{Experimental Input}&~~~~~~~~~  ~~~~~~~~~&\textbf{Theoretical Input}&Eqn.\\
$m_{W'} = 1.9 \; \text{GeV}$ &&  $g_R > g \, s_w / c_w$& \\
$2.4 \; \text{fb} < \sigma_{WZ} <10.2 \; \text{fb}$ && $0 < c_d^2 < 1$ &(\ref{equ:mixing})\\
$46 \; \text{fb} < \sigma_{jj} < 144 \; \text{fb}$ && $\tan \beta > 1$ &\\
&&$1 \leq \rho' \leq 2$ &(\ref{equ:rhoprime})\\
&&$\delta = \infty, 2.5$&(\ref{equ:higgsmassbound})\\
&& $\text{BR}\left(Z' \to \text{SM} \right) = 100 \%, 66 \%$&\\
&&  $\text{BR}\left(W' \to \text{SM} \right) = 100 \%$&
\end{tabular}
\caption{Parameter ranges considered in this analysis.}
\label{tab:params}
\end{table}

In the case that the right handed leptons are embedded in $SU(2)_R$ multiplets, the $Z'$ will be strongly constrained by dilepton resonance searches for sufficiently large $g_R$. ATLAS and CMS have set limits on sequential $Z'$ resonances (which are assumed to have the same couplings to fermions as the SM $Z$ boson) at $\sim 2.8 \; \text{TeV}$ \cite{Aad:2014cka, Khachatryan:2014fba}, and the limit in our model will generically be comparable. Dijet resonance searches are far less constraining for this scenario. There are also important limits on $Z'$ masses and couplings coming from electroweak precision tests, especially those constraining the oblique parameters, four-fermi operators involving at least two leptons, and from measurements of the $Z b \bar{b}$ couplings. In order to assess these constraints we use the formalism and results of \cite{Cacciapaglia:2006pk}. That analysis neglects the constraints coming from four-fermi operators involving right handed quarks as these are generically weaker. However, in the limit of large $g_R$ these might provide important constraints, and so we separately consider the limits on these effective operators derived in \cite{Carpentier:2010ue}. We find that these indirect constraints are always weaker than the ones coming from dilepton resonance searches for the standard lepton embedding.

We also consider the leptophobic case in which the right handed leptons are not charged under $SU(2)_R$. In this scenario the direct constraints coming from dijet and dilepton resonance searches turn out to be comparable and weak. The limits coming from corrections to the oblique parameters then turn out to be the most constraining, which are a consequence of the tree level $Z\text{--}Z'$ mixing given by
\begin{equation}
\sin \theta_{Z Z'} \simeq \frac{g_R}{g} \frac{m_Z^2}{m_{Z'}^2} c_w c_{w'}.
\end{equation}
The constraints from four-fermi operators are weak due to the small coupling of the $Z'$ to leptons, and the corrections to $Z \to b \bar{b}$ are small due to the fact that $b_R$ is mostly an $SU(2)_R$ singlet.

\begin{figure*}
\begin{minipage}[t]{.48\linewidth}
\includegraphics[width=\textwidth]{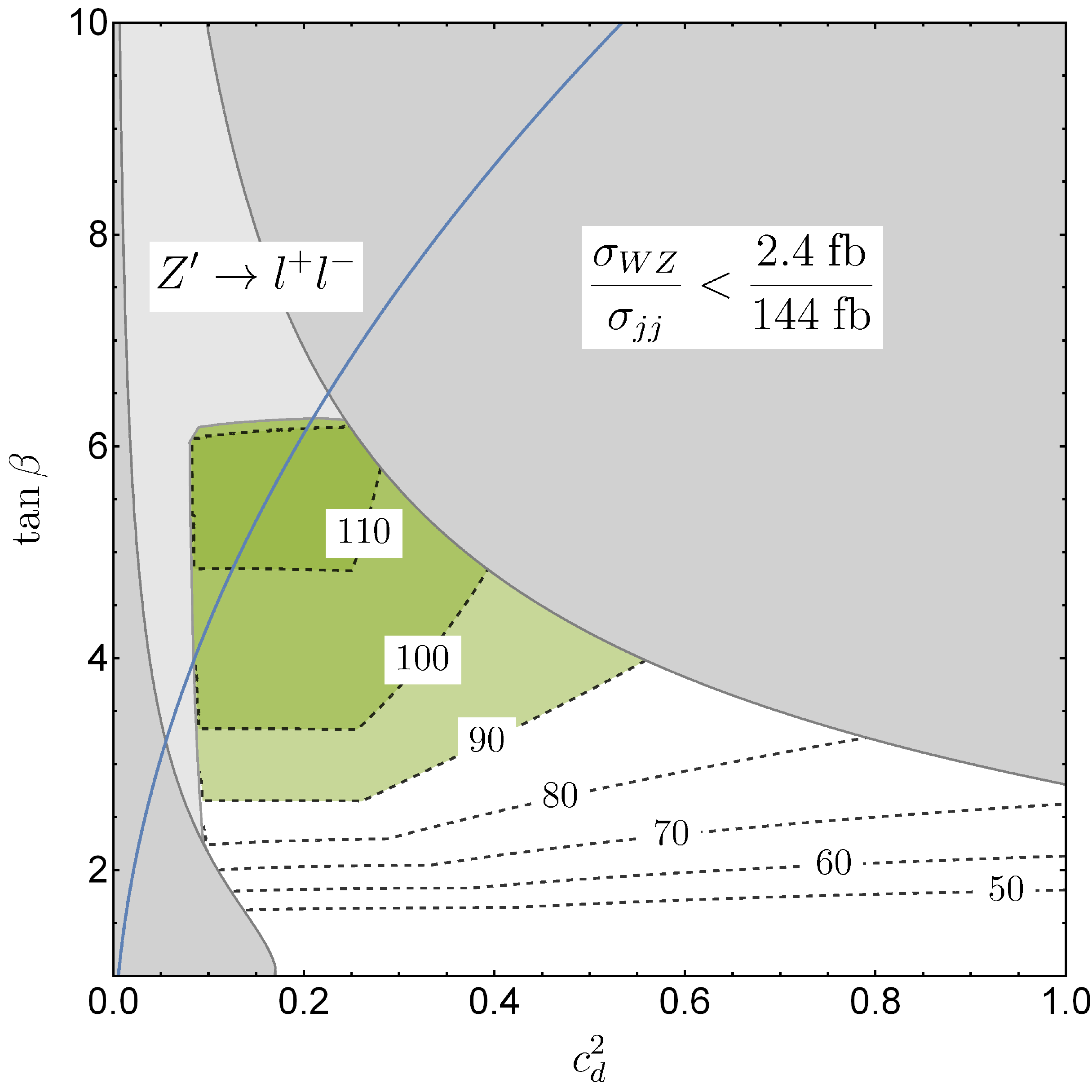}
\end{minipage}\hfill
\begin{minipage}[t]{.48\linewidth}
\includegraphics[width=\textwidth]{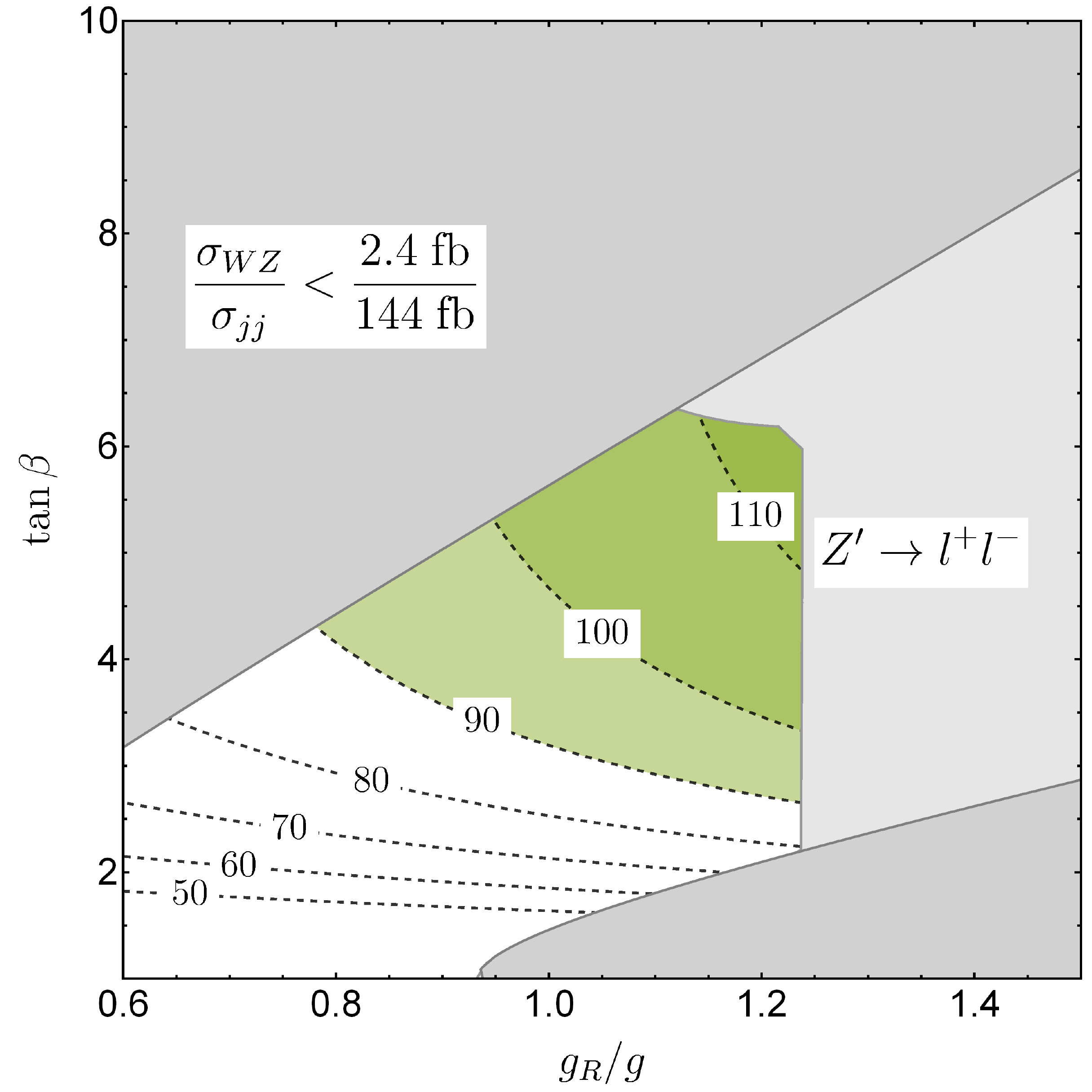}
\end{minipage}
\caption{Maximum tree level Higgs mass from $D$-terms consistent with $W'$ data and limits on $Z' \to \ell^+ \ell^-$, for $\delta = 2.5$ and $\text{BR}(Z' \to \text{SM}) = 100 \%$ with the standard lepton embedding. In each plot we have optimised over all remaining parameters, as explained in the text. Dark grey: Incompatible with the $2 \; \text{TeV}$ anomalies. Light grey: Excluded by $Z' \to \ell \ell$. Contours: Maximum tree level Higgs mass from $D$-terms compatible with the above requirements, in GeV. In the region shaded green, it is possible to exceed the MSSM tree level Higgs mass bound of 91 GeV. The blue line is $t_\beta^2 / c_d^2 = m_c^2 / m_s^2$, near which the charm/strange mass ratio might be explained by the exotic quark mixing.}
\label{fig:normalscans}
\end{figure*}

In \Fig{fig:normalscans} we set $\delta = 2.5$ and $\text{BR}(Z' \to \text{SM}) = 100 \%$ and take the RH leptons to be charged under $SU(2)_R$. In the left plot, we scan the $c_d^2$, $\tan \beta$ plane. In the dark grey region in the top right of the plot, it is not possible to explain the diboson excess without being excluded by dijet resonance searches. This can be seen by noting the ratio between these two widths depends only on $\tan \beta$ and $c_d^2$
\begin{equation}
\frac{\sigma_{WZ}}{\sigma_{jj}} = \frac{\sin^2 (2 \beta)}{24 \, c_d^2}.
\end{equation}
Similarly, the dark grey region in the bottom left of the plot cannot explain the dijet excess without being excluded by the upper limits on the diboson cross section. The remaining region of parameter space is a funnel which can simultaneously explain both excesses. At a generic point in this region, there are a range of values for $g_R$ compatible with the excesses. For small $c_d^2$, $g_R$ is required to be large to generate a sufficiently large $W'$ cross section due to the relationship $\sigma(W') \propto g_R^2 c_d^2$. On the other hand, large $g_R$ reduces the mass splitting between the $Z'$ and the $W'$ and increases the $Z'$ production cross section, while the dominant production channel for this $Z'$ at the LHC is $u_R \bar{u}_R \to Z'$ which is not suppressed by a small mixing angle. The $Z'$ has a significant dilepton branching ratio of $8 \text{--} 18 \%$ and so this region of parameter space is constrained by the dilepton resonance searches. In the light grey region in the top left of the plot, it is not possible to evade the $Z'$ limits while explaining the $W'$ excesses.

\begin{figure*}
\begin{minipage}[t]{.48\linewidth}
\includegraphics[width=\textwidth]{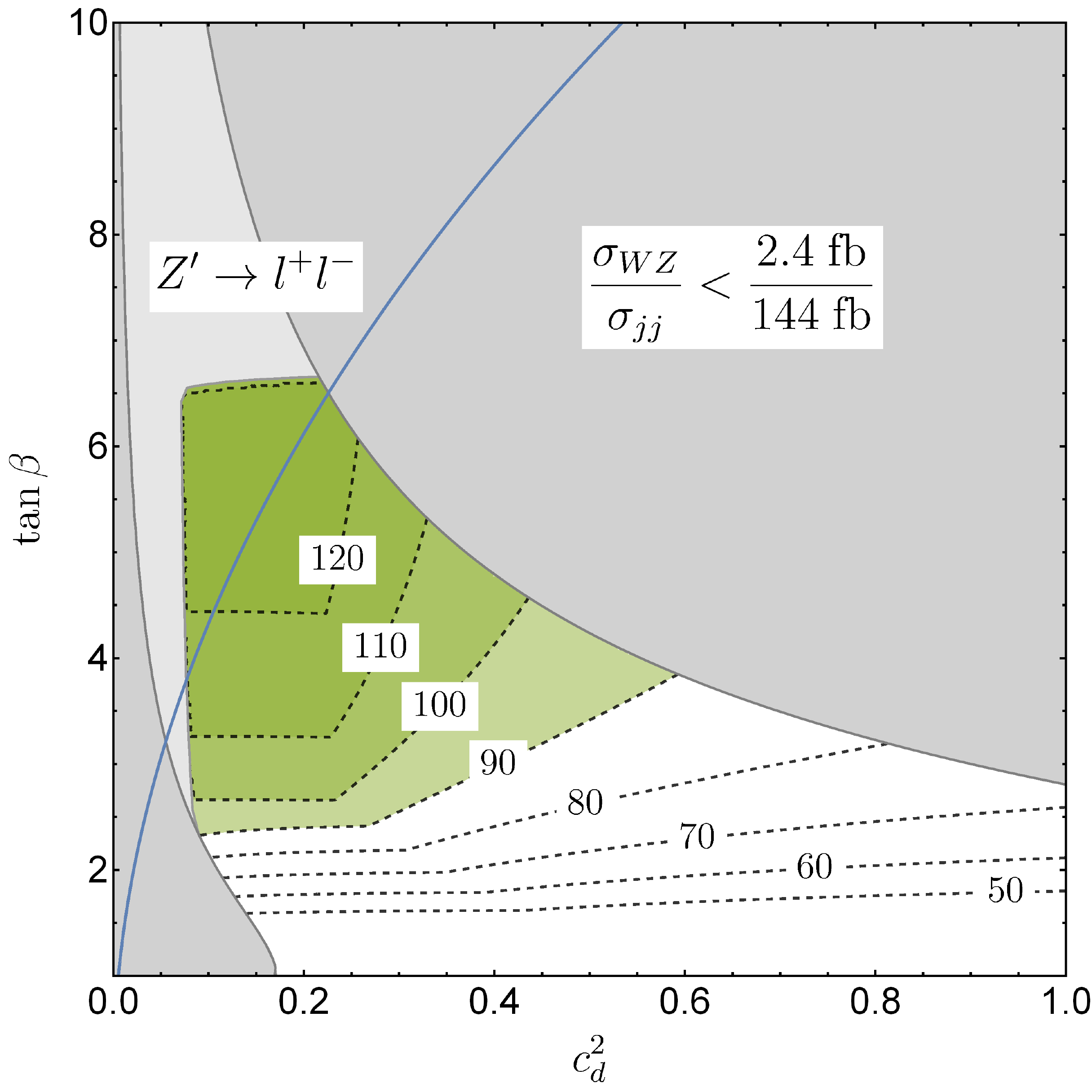}
\end{minipage}\hfill
\begin{minipage}[t]{.48\linewidth}
\includegraphics[width=\textwidth]{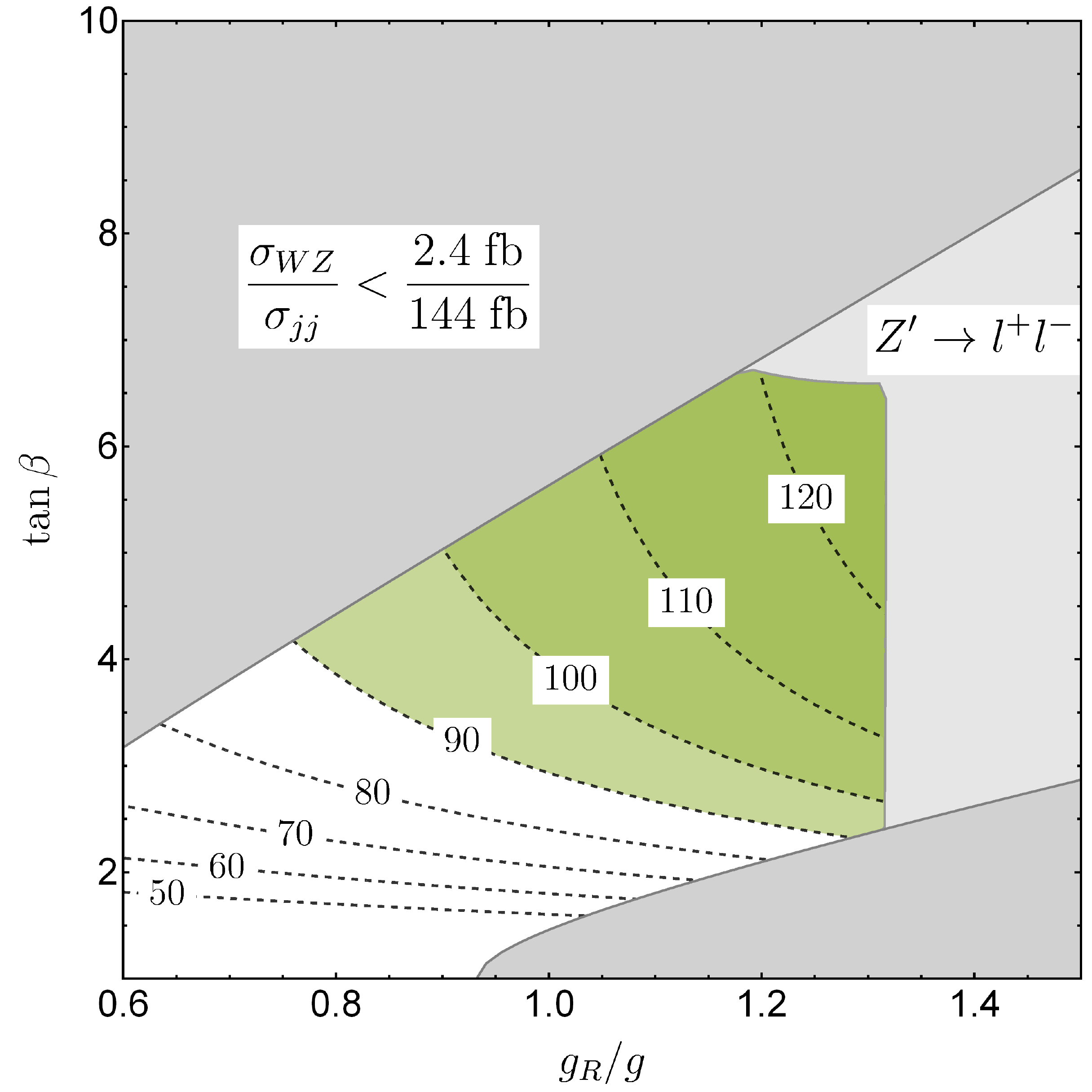}
\end{minipage}
\caption{Maximum tree level Higgs mass from $D$-terms consistent with $W'$ data and limits on $Z' \to \ell^+ \ell^-$, for $\delta = \infty$ and $\text{BR}(Z' \to \text{SM}) = 66 \%$ with the standard lepton embedding. In each plot we have optimised over all remaining parameters, as explained in the text. Dark grey: Incompatible with the $2 \; \text{TeV}$ anomalies. Light grey: Excluded by $Z' \to \ell \ell$. Contours: Maximum tree level Higgs mass from $D$-terms compatible with the above requirements, in GeV. In the region shaded green, it is possible to exceed the MSSM tree level Higgs mass bound of 91 GeV. The blue line is $t_\beta^2 / c_d^2 = m_c^2 / m_s^2$, near which the charm/strange mass ratio might be explained by the exotic quark mixing.}
\label{fig:optimisticscans}
\end{figure*}

\begin{figure*}
\begin{minipage}[t]{.48\linewidth}
\includegraphics[width=\textwidth]{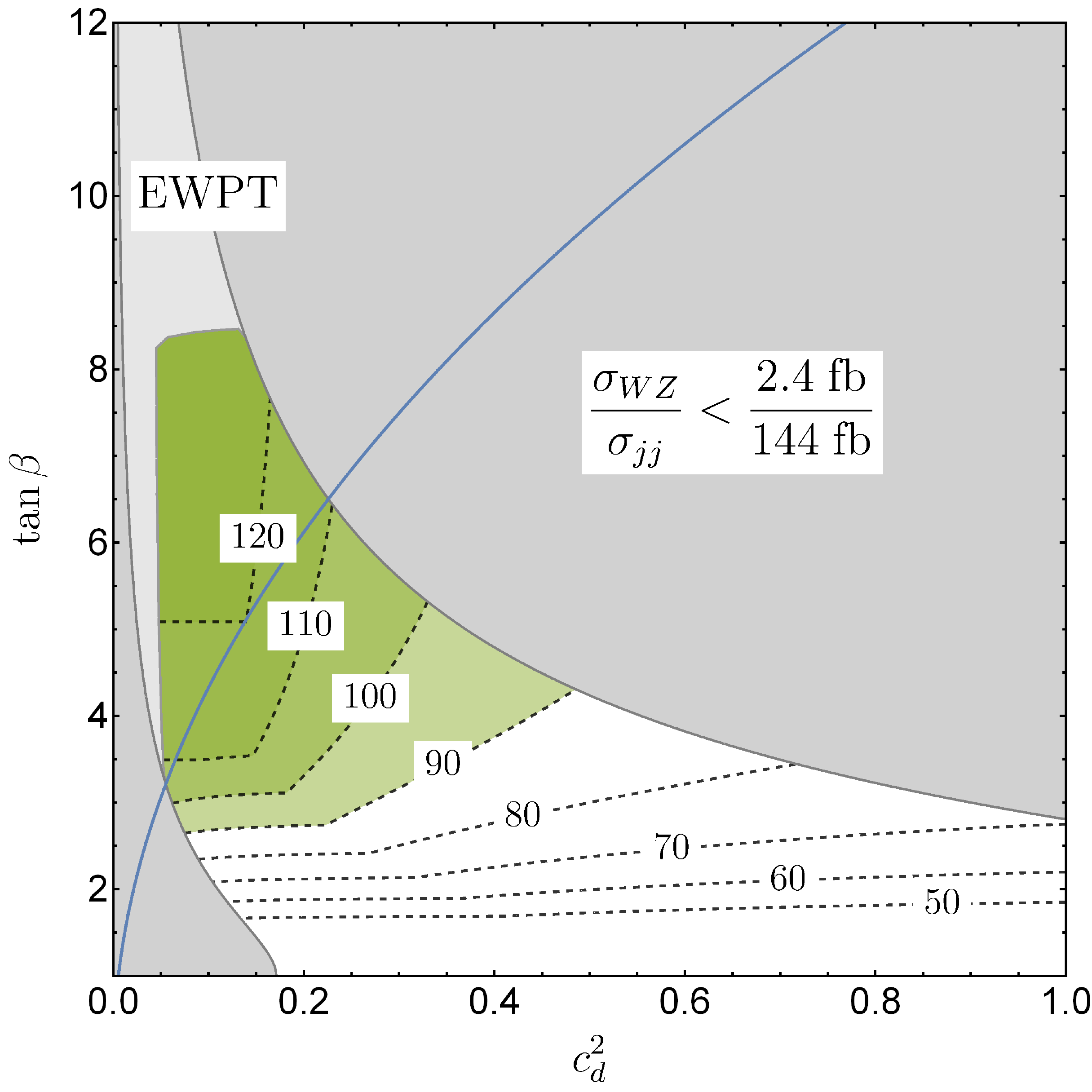}
\end{minipage}\hfill
\begin{minipage}[t]{.48\linewidth}
\includegraphics[width=\textwidth]{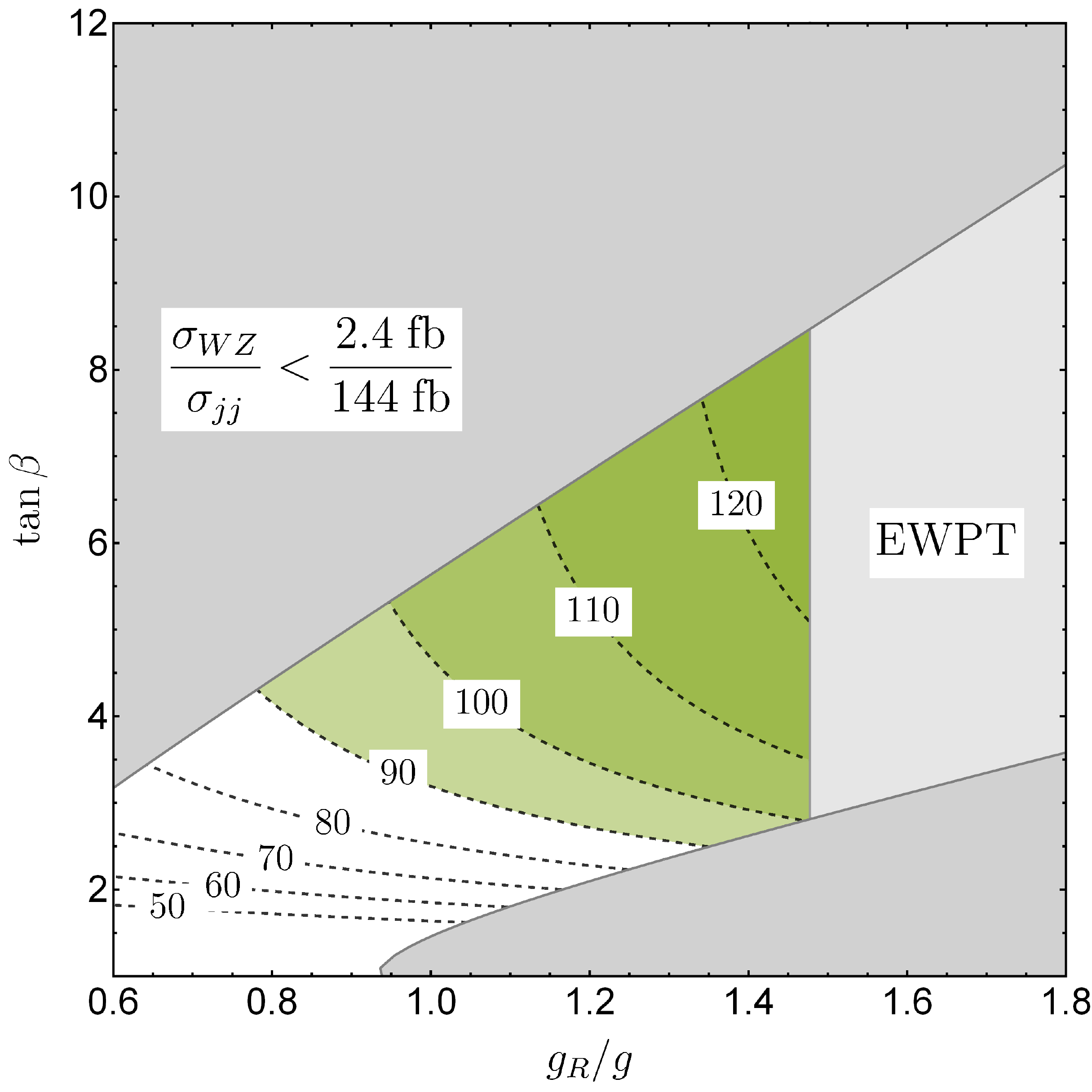}
\end{minipage}
\caption{Maximum tree level Higgs mass from $D$-terms in the leptophobic model consistent with $W'$ data and EWPT, for $\delta = 2.5$. In each plot we have optimised over all remaining parameters, as explained in the text. Dark grey: Incompatible with the $2 \; \text{TeV}$ anomalies. Light grey: Excluded by EWPT. Contours: Maximum tree level Higgs mass from $D$-terms compatible with the above requirements, in GeV. In the region shaded green, it is possible to exceed the MSSM tree level Higgs mass bound of 91 GeV. The blue line is $t_\beta^2 / c_d^2 = m_c^2 / m_s^2$, near which the charm/strange mass ratio might be explained by the exotic quark mixing.}
\label{fig:leptophobicscans}
\end{figure*}

In the surviving region of parameter space we calculate the maximum value of $g_R$ compatible with the constraints and use this to calculate the maximum $D$-term contribution to the Higgs mass, which is shown in GeV by the labelled contours. The region of parameter space compatible with $m_{h, \text{tree}}$ larger than the MSSM tree level bound is highlighted in green. The blue contour highlights the part of parameter space in which the charm/strange mass ratio might be explained by the mixing with the exotic quarks. In the right plot we perform a similar scan in the $g_R$, $\tan \beta$ plane, this time optimising over $c_d^2$. In both plots we have also optimised over $\rho'$ and over the parton luminosities within the $1\sigma$ uncertainties calculated from the NNPDF ensemble, assuming that the uncertainties on $W'$ and $Z'$ production are completely correlated. In practise, this means setting $\rho' = 2$ and using the lower prediction for the parton luminosities, except for a narrow band at large $\tan \beta$ where higher estimates are preferred. In \Fig{fig:optimisticscans} we perform a similar scan for $\delta = \infty$ and $\text{BR}(Z' \to \text{SM}) = 66 \%$. This would be the enhancement in the $Z'$ width if, for example, every SM fermion had a light SUSY partner. In \Fig{fig:leptophobicscans} we consider a leptophobic scenario with $\delta = 2.5$ and $\text{BR}(Z' \to \text{SM}) = 100 \%$. This time the paramer space is constrained by indirect constraints on the $Z'$ in the regions labelled `EWPT'. In all other respects the procedure is the same as for the previous plots.

We see that there is a region of parameter space with $0.1 \lesssim c_d^2 \lesssim 0.5$, $2.5 \lesssim \tan \beta \lesssim 6$ and $0.8 \lesssim g_R / g \lesssim 1.2$ with a $D$-term contribution to the Higgs mass at least as large as the MSSM tree level bound consistent with requirement of explaining the $2 \; \text{TeV}$ and evading $Z'$ limits. Fine tuning considerations are model dependent, but a tree level Higgs mass of $110 \; \text{GeV}$ is compatible with $\delta = 2.5$ which need not be associated with significant fine tuning. Allowing the $Z'$ to have a significant branching fraction into non SM states allows for a broader region of parameter space to explain the excess, as illustrated in \Fig{fig:optimisticscans}, though the main impact on the Higgs mass in this scan (which may exceed 120 GeV) comes from taking the decoupling limit $\delta \to \infty$ which would come with a significant fine tuning penalty. Due to the weaker $Z'$ bounds, the leptophobic model allows for the greatest $D$-term Higgs quartic as larger values of $g_R$ and $\tan \beta$ are permitted. A tree level Higgs mass of 120 GeV is possible in this model with $\delta = 2.5$. Note that the line $c_d^2 = 1$ which corresponds to the model without the exotic quarks cannot accomadate a tree level Higgs mass larger than 70~GeV while explaining the excess.

\subsection{Implications for the $Z'$ and stops}

\begin{figure*}[t]
\begin{minipage}[t]{.49\linewidth}
\includegraphics[width=\textwidth]{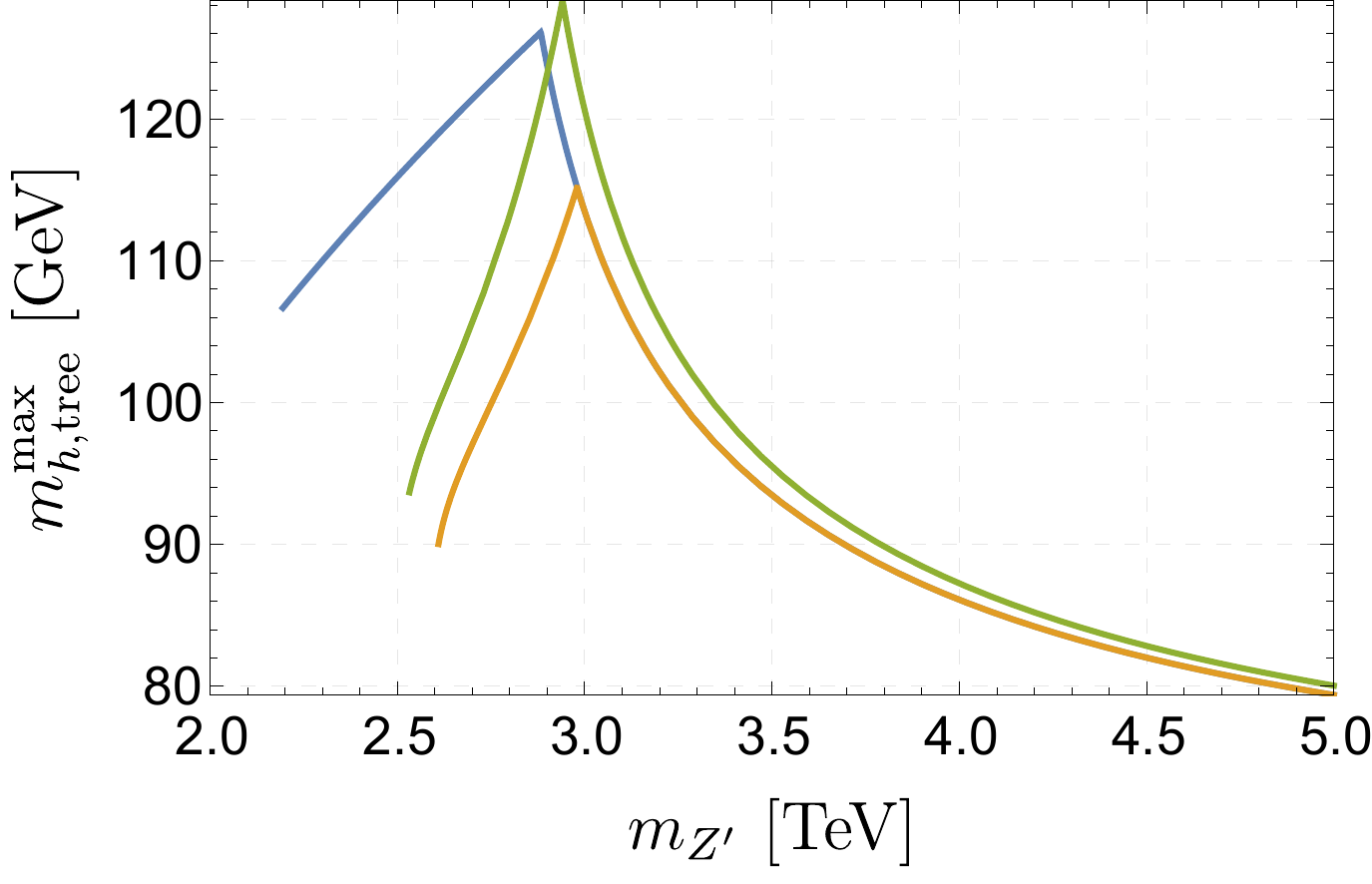}
\end{minipage}\hfill
\begin{minipage}[t]{.48\linewidth}
\includegraphics[width=\textwidth]{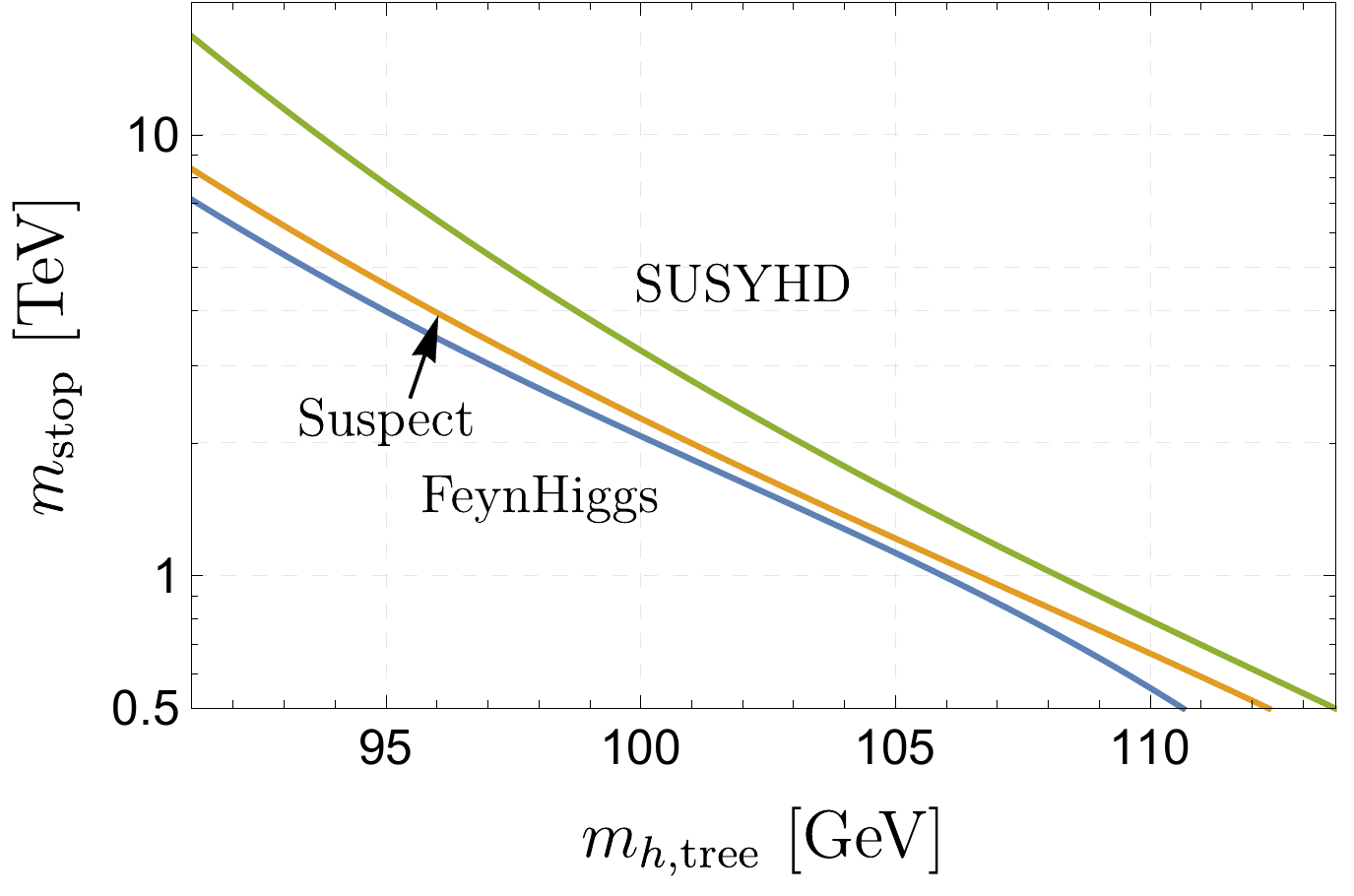}
\end{minipage}
\caption{Left: Maximum $D$-term contribution to the tree level Higgs mass for a given value of $m_{Z'}$ in three scenarios. Orange: $\delta = 2.5$, $\text{BR}(Z' \to \text{SM}) = 100 \%$. Green: $\delta = \infty$, $\text{BR}(Z' \to \text{SM}) = 66 \%$. Blue: Leptophobic, $\delta = 2.5$. Each line stops at small $m_{Z'}$ at the limit from direct searches for dilepton resonances, or electroweak precision constraints in the case of the leptophobic model. Right: Stop mass required to raise the Higgs mass to 125 GeV without left-right stop mixing.}
\label{fig:maxhiggsmass}
\end{figure*}

Due to the constraints on $g_R$, there is a close relation between the $Z'$ mass and the possible enhancement to the Higgs mass. In the left of \Fig{fig:maxhiggsmass} we plot the maximum possible tree level Higgs mass compatible with all constraints as a function of $m_{Z'}$ in each of the three scenarios described above. For large $m_{Z'}$ the size of $g_R$ is limited by \Equ{equ:rhoprime} and the requirement $\rho' \leq 2$, and this is the main constraint on the Higgs mass for $m_{Z'} \gtrsim 3 \; \text{TeV}$. Converseley, small $m_{Z'}$ corresponds to larger values of $g_R$. In this case, the main constraint on the Higgs mass are the direct or indirect limits on the $Z'$. The kinks represent the transition between these scenarios. We see that the requirement $m_{h, \text{tree}} > 100 \; \text{GeV}$ can be satisfied only for $2.6 \; \text{TeV} \leq m_{Z'} \leq 3.3 \; \text{TeV}$, assuming the right handed leptons have $SU(2)_R$ charge. The Higgs mass is maximized for $m_{Z'} \simeq 2.95 \; \text{TeV}$. This result is especially interesting in light of the anomalous $2.9 \; \text{TeV}$ dilepton event observed by the CMS experiment with $65 \; \text{pb}^{-1}$ of integrated luminosity \cite{CMS-DP-2015-039}. In the case of a leptophobic $Z'$, its mass might be as low as $2.2 \; \text{TeV}$ while still permitting a large $D$-term contribution to the Higgs mass.

We now turn to a brief discussion of the radiative corrections to the Higgs mass. In the MSSM, the Higgs mass might be raised to 125 GeV by a large stop mass, but this loop contribution grows more slowly than $\log(m_{\tilde{t}}^2 / m_t^2)$, requiring $m_{\tilde{t}} \sim 10\; \text{TeV}$ in the absence of large mixing between the LH and RH stops. This might be reduced to $\sim (2\text{ -- }5) \; \text{TeV}$ for sufficiently large mixing in the stop sector. A comparison of results using diagrammatic and effective field theory techniques can be found in \cite{Vega:2015fna}, which compares the codes SUSYHD \cite{Vega:2015fna}, FeynHiggs \cite{Heinemeyer:1998yj}, and SuSpect \cite{Djouadi:2002ze}. In our model the requirements on the stop sector will be significantly relaxed due to the increased tree level contribution to the Higgs mass. There will also be additional radiative corrections due to the new (s)quark states in the third generation, but these will not be log enhanced if there is not a large splitting between the exotic quark and squark masses and so are expected to be subdominant compared to the stop contributions. In lieu of a complete calculation of the radiative corrections in this model, we use the following approximation to estimate the relaxed requirements on the stop sector. We consider the MSSM contribution in the limit of no left-right mixing and large $\tan \beta$ and define the function
\begin{equation}
 \Delta(m_{\tilde{t}}^2) = {m_h^2}_{\text{(MSSM)}}(m_{\tilde{t}}^2) - m_Z^2.
\end{equation}
This function can be taken from the SUSYHD, FeynHiggs, and SuSpect calculations. We then estimate the Higgs mass with the new tree level contributions as:
\begin{equation}
m_h^2(m_{\tilde{t}}^2) = m_{h, \text{tree}}^2 +  \Delta(m_{\tilde{t}}^2).
\end{equation}
This approximation neglects additional wavefunction renormalization effects due to the enhanced Higgs quartic, and threshold effects from the exotic states. In \Fig{fig:maxhiggsmass} right we plot the stop mass required to achieve a Higgs mass of 125 GeV using this approximation. We see that stops may be lighter than 1~TeV in this model, alleviating their contribution to the little hierarchy problem of the MSSM.

\section{\label{sec:flavour}Flavour constraints}

There have been numerous studies of flavour constraints on generic LRS models \cite{Ecker:1985vv, Nishiura:1990hh, Deshpande:1990ip, Zhang:2007da, Maiezza:2010ic} and on models with vector-like down-type quarks \cite{Bjorken:2002vt, Golowich:2007ka}. However, certain features of our model prevent direct application of the existing constraints, and hence necessitates a separate analysis. First, most constraints on LRS do not consider the effects of large mixings with vector-like quarks in the RH sector, which reduces the coupling of the physical light quarks to the RH gauge bosons. Second, the strongest constraints on most vector-like quark extensions to the SM typically comes from tree-level $Z$ FCNCs involving LH quarks due to violation of the Glashow-Weinberg-Paschos condition \cite{Glashow:1976nt,Paschos:1976ay}. However, this constraint is much weaker in our model since terms of the form $H_d Q_L D^{\prime c}$ are now forbidden by the RH gauge symmetry, as a result of which there is an additional Yukawa suppression in the mixing between $d'_L$ and $D'$. Besides the above constraints, we also have contributions to FCNCs that involve the superpartners, in particular new box diagrams involving gluinos and the exotic squarks. On the other hand, these depend on parameters such as soft squark masses which are not closely related to the phenomenology discussed in the previous sections. There is also the possibility of cancellations between gauge boson and supersymmetric diagrams as suggested in \cite{Zhang:2007qma}.

Since the complete analysis of all flavour constraints on the model is a rather formidable task, we have restricted our attention to mainly tree-level and a small subset of one-loop $|\Delta F|=2$ FCNC processes that are directly related to the new quarks. We postpone a more complete analysis, including CP violation and other FCNC processes such as $b\rightarrow s\gamma$ to future work. We find that the strongest constraints come from tree-level $Z'$ FCNCs involving the RH quarks, which we discuss in this section. Details of the conventions used and constraints from other FCNC diagrams are presented in the appendix.

\subsection{Tree-level $Z'$ FCNCs}

The interaction basis $d^{\prime c}_R$ and $D^{\prime c}$ can be written in terms of the mass basis as
\begin{equation}
\begin{aligned}
d^{\prime c}_R &= (c_R U^d_R)^* d_R^c + \mbox{ terms involving $D^c$},\\
D^{\prime c} &= (s_R U^d_R)^* d_R^c + \mbox{ terms involving $D^c$}.
\end{aligned}
\end{equation}
Here, $c_R$ and $s_R$ are matrices that describe the mixing between the $d^{\prime c}_R$ and $D^{\prime c}$ as discussed in \Sec{sec:exotic}, except that we no longer assume $c_R$ to be a diagonal matrix with elements $c_d$ and $c_b$. $U^d_R$ is the RH unitary transformation that diagonalises the light down-type mass matrix obtained from the pre-diagonalisation with $c_R$ and $s_R$. For convenience, we also define the RH equivalent of the CKM matrix
\begin{equation}
V_{\text{CKM}}^R \equiv c_R U_d^R.
\end{equation}
Further details of the definitions above can be found in the appendix.

Since $d^{\prime c}_R$ and $D^{\prime c}$ couple to $Z'$ differently, the $Z'$-coupling to the mass basis $d_R$ is non-universal and given by the matrix $\mathcal{C}^{\text{tree}}_{Z-d_L}$, defined as
\begin{equation}
\begin{aligned}
\mathcal{L} &\supset  Z^{\prime \mu} \overline{d_R^c} \mathcal{C}^{\text{tree}}_{Z'-d_R^c} \bar{\sigma}_\mu d_R^c,\\
\mathcal{C}^{\text{tree}}_{Z'-d_R^c} &= \frac{g_R}{c_{w'}} \left[ \frac{1}{2} V_{\text{CKM}}^{R\dagger} V_{\text{CKM}}^R -\frac{1}{3} s_{w'}^2\right]^T.
\end{aligned}
\end{equation}
We now consider $|\Delta F| = 2$ processes, in particular $K-\bar{K}$ mixing. While there is a large mass suppression from $m_{Z'}$ in the propagator, if we simply regard $c_R$ as a completely generic matrix of order $\mathcal{O}(c_d)$, the contribution to $\Delta m_K$ ends up being much larger than the experimental constraints. Instead, we require that
\begin{equation}
\left( \frac{g_R/g}{1.2} \right) \left( \frac{0.9}{c_{w'}} \right) \left( \frac{3 \mbox{ TeV}}{m_{Z'}}\right)^2 \left( \frac{(V_{\text{CKM}}^{R\dagger} V_{\text{CKM}}^R)_{12}}{0.2} \right)^2 \lesssim 0.001
\end{equation}
in order to satisfy bounds on $\Delta m_K$ \cite{Isidori:2010kg}. (Note that $g_R/g$ and $c_{w'}$ should not be regarded as independent parameters.) In other words, the 12 elements of $V_{\text{CKM}}^{R\dagger} V_{\text{CKM}}^R$ should be much smaller than typical values of $\mathcal{O}(c_d^2)$.

To achieve a small $(V_{\text{CKM}}^{R\dagger} V_{\text{CKM}}^R)_{12}$, one possibility is to consider an analogue of the Glashow-Iliopoulos-Maiani (GIM) mechanism. We recall that $V_{\text{CKM}}^{R\dagger} V_{\text{CKM}}^R = (U_R^d)^\dagger c_R^\dagger c_R U_R^d$, and that $U_R^d$ is unitary. Therefore, should $c_R^\dagger c_R$ be proportional to the identity matrix, the same will be true for $V_{\text{CKM}}^{R\dagger} V_{\text{CKM}}^R$ so off-diagonal elements become zero. One could impose an approximate $U(3)$ flavour symmetry such that all the couplings are universal, in which case $c_R$ is itself proportional to the identity. However, this is inconsistent with the down-type mass spectrum which requires that the third diagonal element $c_b$ be somewhat smaller than the first two elements $c_d$. Instead, we impose an approximate $U(2)$ symmetry for the first two generations, and further require that the mixings with the third generation be small. This ensures that $c_R$ remain approximately diagonal, while also suppressing the 31 and 32 elements of $U_R^d$. The suppression is required since the GIM cancellation is now incomplete.

To quantify the constraints on $z$ and $M$, we work in a $D'-D^{\prime c}$ basis such that $M$ is diagonal. We then parameterise $z$ as $Uz^{\text{diag}}V$, where $U$ and $V$ are arbitrary unitary matrices. For simplicity, we assume the 12 rotation angles in both matrices be of the same order $\mathcal{O}(\theta_{12})$, and the 13 and 23 rotation angles be of order $\mathcal{O}(\theta_3)$. We also define a parameter $\delta$ that quantifies the breaking of the universality in the first two generations, i.e. we expect that $M_{22}/M_{11}$ and $(z^{\text{diag}})_{22}/(z^{\text{diag}})_{11}$ are both $1+\mathcal{O}(\delta)$. In view of the requirements on $c_R$, we expect a strong constraint on $\theta_{3}$, and a possibly weaker constraint on $\theta_{12}$ that depends on $\delta$.

\begin{figure}[t]
\centering
\includegraphics[width=0.5\linewidth]{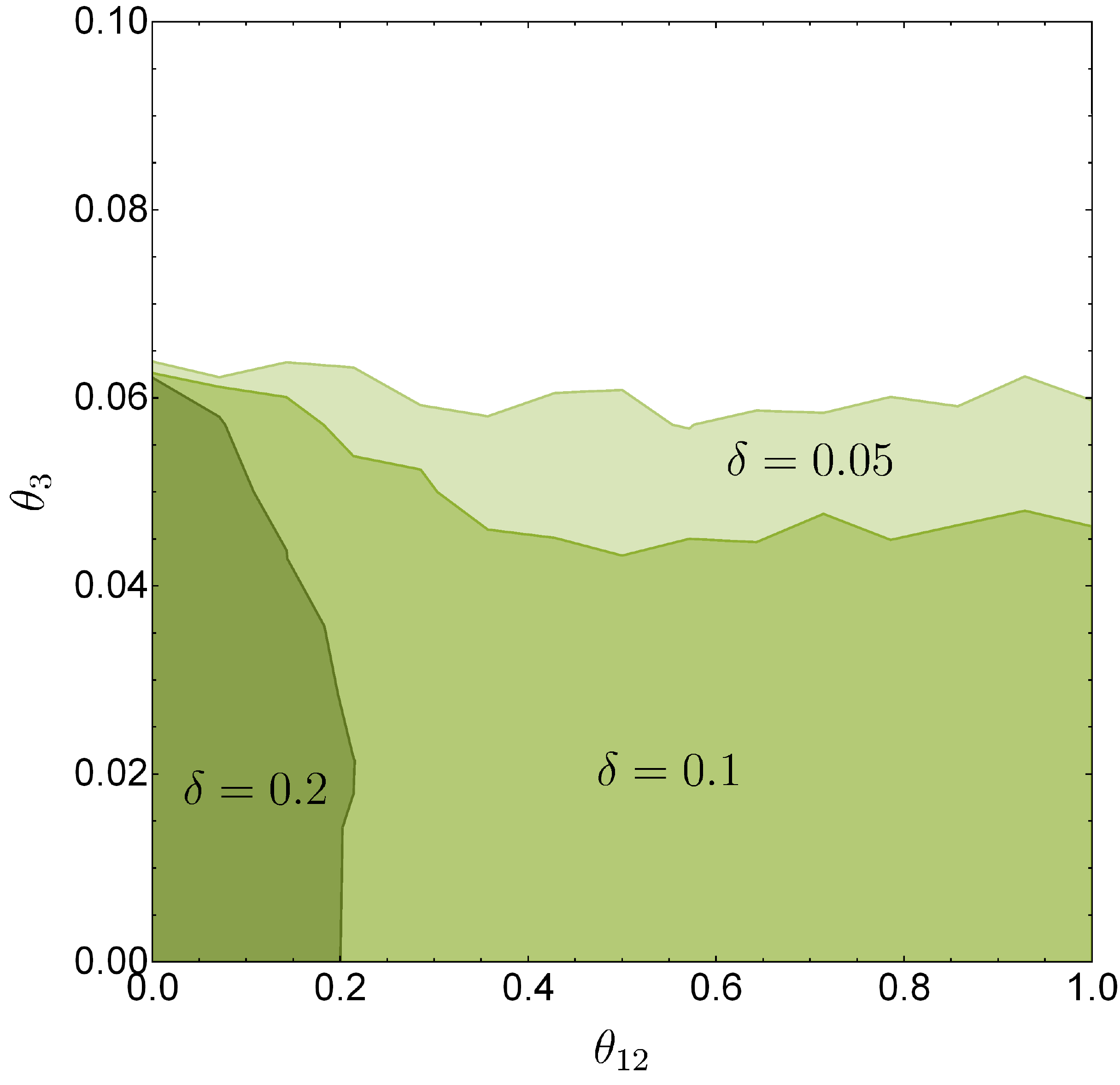}
\caption{Regions in parameter space allowed by the tree-level $Z'$ FCNC constraints. $\theta_{12}$ gives the characteristic size of 12 alignment angles, while $\theta_{3}$ does so for the 13 and 23 angles. $\delta$ quantifies the breaking of the universality in the first two generations.} \label{fig:zprime-scan}
\end{figure}

\Fig{fig:zprime-scan} shows regions of $\theta_{12}$ and $\theta_3$ for different $\delta$ allowed by the tree-level $Z'$ FCNC constraint. For each choice of the three parameters $\theta_{12}$, $\theta_{13}$ and $\delta$, 1000 sets of mixing angles, $M$ and $z^{\text{diag}}$ are then randomly generated with characteristic sizes specified by the parameters. A parameter choice is ``allowed'' if at least half of the corresponding 1000 random sets are found to satisfy the $Z'$ constraints. We see from the plot that $\theta_3$ should be at most $\mathcal{O}(0.05 \mbox{ rad})$ which is comparable to $(V_{\text{CKM}}^L)_{13}$ and $(V_{\text{CKM}}^L)_{23}$, suggesting an alignment similar to what is already in the SM. Meanwhile, the constraints on $\theta_{12}$ are as expected much weaker should the extent of universality breaking be small. For example, a $5\%$ breaking will allow for a alignment angle of more than $1 \mbox{ rad}$.

\section{\label{sec:conclusions}Conclusions}

We have explored the possibility that an $SU(2)_R$ gauge extention of the MSSM which is compatible with an explanation of the recent diboson, $eejj$ and dijet resonance excesses in terms of a 1.9 TeV $W_R$ might also give rise to a significant non-decoupling $D$-term enhancement to the Higgs mass. The inferred diboson cross section is relatively large compared to the dijet cross section, which requires $\tan \beta \simeq 1$ in minimal models. Furthermore, the total $W'$ cross section implies $g_R < 0.8 \, g$ in minimal models. Both of these features are not compatible with large $D$-terms for the Higgs which requires large $\tan \beta$ and large $g_R$, and the small value for $\tan \beta$ is also problematic for the top/bottom mass ratio.

We have therefore been lead to an extended model which also includes a charge $-1/3$ vector-like quark for each generation which mixes significantly with the $SU(2)_R$ doublets after that gauge symmetry is broken. We have assumed that the mixing angle is approximately universal for the first two generations, but may be different for the third. For the third generation, this means that $m_b/m_t$ is suppressed by both $t_\beta^{-1}$ and the cosine of the mixing angle, $c_b$. More importantly, the mixing angle suppresses the $W_R$ couplings to the SM quarks, enhancing the diboson to dijet signal cross section ratio by a factor $c_d^{-2}$, and suppresses the $W_R$ production cross section by a factor $c_d^2$. This allows the excesses to be fit with larger $g_R$ and $\tan \beta$, compatible with the Higgs mass requirement. An additional key difference compared with previous discussions is the suppression of the branching fraction of the $W'$ into $tb$ which is a consequence of the mechanism we have chosen for achieving the mass ratio $m_t/m_b$. Without this suppression, the absence of a signature in this channel so far is a leading constraint on $g_R$.

The additional quark fields raise many questions related to flavour physics, and we have addressed some of these questions in this paper. We have argued that the strongest constraints on the flavour structure of this new sector will come from FCNCs induced at tree level by the flavour-violating couplings of the $Z'$. Nonetheless, an approximate universality among the first two generations combined with an alignment of the mixing angles with the third generation comparable in size with that already present in the SM CKM matrix allows us to evade those constraints. Additional contributions to FCNC observables in the quark and lepton sectors are expected to come from the extended Higgs sector and from squark and gaugino loops. Furthermore, we have not yet provided a complete account of the generation of the full flavour structure of the quark sector in the SM. On the other hand, we have found that the region of parameter space which maximises the Higgs mass is also compatible with the naive expectation for the charm/strange mass ratio if this is purely a consequence of $\tan \beta$ and the mixing with the exotic quarks. We leave a complete analysis of the flavour structure of this model to future work. While our analysis places no direct constraints on the masses of the exotic quarks, it is possible that they are sufficiently light to be directly pair-produced and observed at run 2 of the LHC. A natural expectation is that they will decay into $D \to j Z$ with a significant branching fraction via the $Z\text{--}Z'$ mixing, which would provide an opportunity to directly measure their mass from the invariant mass of the $j$ and $Z$.

The essential result of our analysis is that we have identified a region of parameter space in a model with right handed leptons charged under $SU(2)_R$ with $m_{h, \text{tree}} > 100 \; \text{GeV}$ for $0.1 \lesssim c_d^2 \lesssim 0.4$, $3 \lesssim \tan \beta \lesssim 6$, and $1 \lesssim g_R / g \lesssim 1.2$ without imposing an irreducible fine tuning. This region is broadened by relaxing the assumption $\text{BR}\left(Z' \to \text{SM}\right) = 100 \%$ and by taking the extreme decoupling limit for the $D$-terms. The leptophobic scenario in which the right handed leptons are not embedded in $SU(2)_R$ multiplets is also more weakly constrained and allows for a larger contribution to the Higgs mass.

A key finding of this analysis is that the possible $D$-term enhancement of the Higgs mass is closely related to the $Z'$ mass. A light $Z'$ is favoured for raising the Higgs mass, as this corresponds to larger $g_R$. On the other hand the $Z'$ is quite constrained by dilepton resonance searches from LHC run 1 or electroweak precision measurements for $m_{Z'} \lesssim 3 \; \text{TeV}$. We find that with the standard lepton embedding, the range $2.6 \; \text{TeV} < m_{Z'} < 3.3 \; \text{TeV}$ is compatible with $m_{h, \text{tree}} > 100 \; \text{GeV}$, while the Higgs mass bound is optimised for $m_{Z'} \simeq 2.95 \; \text{TeV}$. This scenario should result in a clear dilepton resonance at run 2 of the LHC. On the other hand, the leptophobic scenario is compatible with large contributions to the Higgs mass and is not excluded for $m_{Z'} > 2.2 \; \text{TeV}$. This $Z'$ is more challenging to discover at the LHC. Looking forward we should be paying close attention to any hints of a 2~TeV resonance in the new data from the LHC, but we should bear in mind that the broader and potentially quite significant implications of such a resonance might depend sensitively on the results of searches for related particles like a $Z'$, vector-like quarks and leptons, massive neutrinos, etc.

\begin{acknowledgements}
We would like to thank Maxim Perelstein for invaluable guidance and encouragement in the early stages of this work, and for pointing out interesting collider signatures of the exotic quarks. We would also like to thank MP, Csaba Csaki, Yuval Grossman and Marco Farina for helpful discussions and comments on the final manuscript.
\end{acknowledgements}

\appendix
\section{\label{sec:couplings}$W'$ and $Z'$ couplings and partial widths}

The partial widths for the $W'$ are taken as
\begin{align}
\Gamma(W' \to jj) &= \frac{g_R^2 c_d^2}{8 \pi}m_{W'},\\
\Gamma(W' \to tb) &= \frac{g_R^2 c_b^2}{16 \pi}m_{W'}\left(1 + \mathcal{O}\left(\frac{m_t^2}{m_{W'}^2}\right)\right),\\
\Gamma(W' \to WZ) &= \frac{g_R^2}{192 \pi} m_{W'} \sin^2 2 \beta \left(1 + \mathcal{O}\left(\frac{m_W^2}{m_{W'}^2}\right)\right),\\
\Gamma(W' \to Wh) &= \frac{g_R^2}{192 \pi} m_{W'} \cos^2(\alpha + \beta) \left(1 + \mathcal{O}\left(\frac{m_W^2}{m_{W'}^2}\right)\right).
\end{align}
We take the decoupling or alignment limit for the Higgs, with $\alpha = \beta + \pi/2$. Calculating the $W'$ production cross section requires the the coupling $\mathcal{L} \supset g_{W'u_R d_R} W^-  u_R d_R^c + \text{h.c.}$ which is given by
\begin{equation}
g_{W'u_R d_R} = g_R c_d.
\end{equation}

The $Z'$ couplings to SM fermions, defined by $\mathcal{L} \supset g_{Z' f f'} Z' f f'$, are given in the flavour conserving limit by
\begin{equation}
g_{Z' f \bar{f}} = \frac{g_R}{c_{w'}}\left(c_d^2 T^3_R - s_{w'}^2 Y \right)
\end{equation}
where $c_d$ is the cosine of the mixing angle of the SM quark into an $SU(2)_R$ state. In particular, the couplings are
\begin{align}
g_{Z' u_R^c u_R^c} &= \frac{g_R}{c_{w'}} \left(-\frac{1}{2} + \frac{2}{3} s_{w'}^2 \right) &g_{Z' u_L u_L} &= -\frac{1}{3}\frac{s_{w'}^2}{c_{w'}}g_R\\
g_{Z' d_R^c d_R^c} &= \frac{g_R}{c_{w'}} \left(\frac{c_d^2}{2} - \frac{1}{3}s_{w'}^2 \right) &g_{Z' d_L d_L} &=-\frac{1}{3}\frac{s_{w'}^2}{c_{w'}}g_R\\
g_{Z' \ell_R^c \ell_R^c} &= \frac{g_R}{c_{w'}} \left(\frac{1}{2} - s_{w'}^2 \right) &g_{Z' \ell_L \ell_L} &=-\frac{1}{2}\frac{s_{w'}^2}{c_{w'}}g_R\\
g_{Z' \nu_R^c \nu_R^c} &=- \frac{1}{2} \frac{g_R}{c_{w'}} &g_{Z' \nu_L \nu_L} &=-\frac{1}{2}\frac{s_{w'}^2}{c_{w'}}g_R
\end{align}
The partial width to fermions is then given by (up to corrections of order $m_f^2 / m_{Z'}^2$)
\begin{equation}
\Gamma(Z' \to f \bar{f}) = \frac{N_c}{24 \pi}m_{Z'}g_{Z' f \bar{f}}^2.
\end{equation}
The partial widths into SM bosons, again up to corrections suppressed by $m_{Z'}^2$, are given by
\begin{align}
\Gamma(Z' \to WW) &= \frac{g_R^2}{192 \pi} m_{Z'} c_{w'} \sin^2 2 \beta\\
\Gamma(Z' \to Zh) &= \frac{g_R^2}{192 \pi} m_{Z'} c_{w'} \cos^2\left( \alpha + \beta \right).
\end{align}
The width into $W W'$ is suppressed by $m_W^2 / m_W'^2$ compared to those above.

\begin{figure}
\centering
\includegraphics[width=0.5\linewidth]{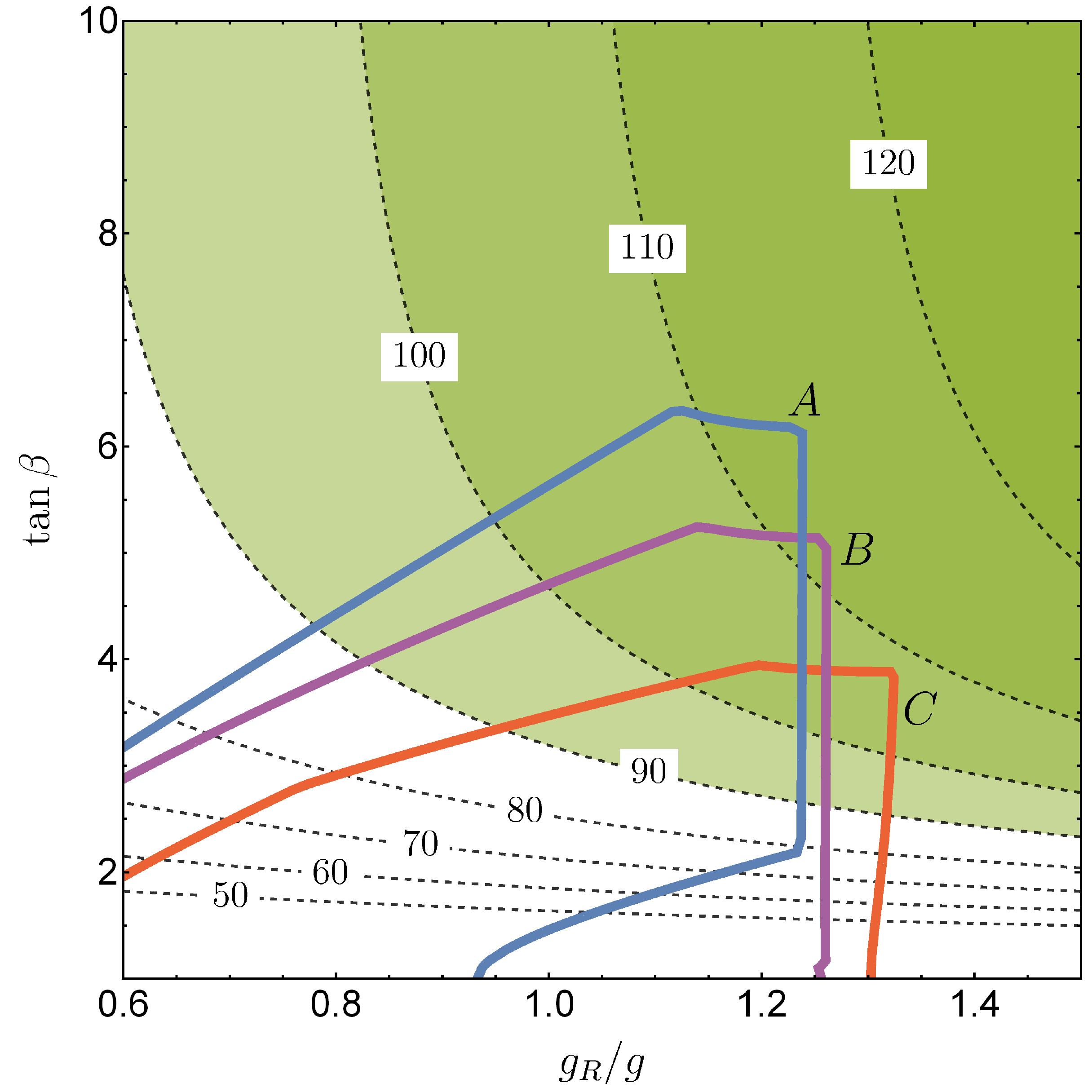}
\caption{Allowed parameter space when considering additional BSM decays for $W'$ and $Z'$. In each case, the region above and to the right of the coloured line is excluded. Case A: Only decays to SM states. Case B: Include decays of the $W'$ and $Z'$ involving a single light generation of RH neutrino with $m_{\nu_R} \ll m_W'$. Case C: Inlcude decays of the $W'$ and $Z'$ to the first two generations of exotic quark, with $m_D \ll m_{W'}$. In all other respects the plot is generated as described in the caption to \Fig{fig:normalscans} and the related text in \Sec{sec:results}.}
\label{fig:extraBRplot}
\end{figure}

In \Fig{fig:extraBRplot} we explore the effect of allowing the $W'$ and $Z'$ to decay into right handed neutrinos or first and second generation vector-like quark. In each case we assume that the new particles are very light, and neglect any kinematic suppression from their masses. There are two main effects at work. Firstly, the additional channels dilute the $W'$ diboson signature, requiring a larger value of $\sin^2 2 \beta$ and a smaller value of $\tan \beta$. Secondly, these channels also dilute the constraining $Z' \to \ell \ell$ signature, allowing for larger values of $g_R$. The net effect is a small reduction in the allowed size of the tree level Higgs mass from $D$-terms.

\section{\label{sec:tuning}Non-Decoupling D-terms and Fine Tuning}

We consider a simple model to illustrate the main features in the relationship between the decoupling parameter $\delta$ and the parameters of the $SU(2)_R$ breaking potential. Suppose that $v_D^2 \ll v_T^2$ so that we have a triplet breaking scenario, and the triplet has the superpotential
\begin{equation}
W = \lambda S \left( \Delta \xbar{\Delta} - f^2 \right),
\end{equation}
where we have introduced a singlet $S$. We also introduce soft masses
\begin{equation}
V_\text{soft} = m_S^2 S^\dagger S + m_\Delta^2 \Delta^\dagger \Delta + m_{\bar{\Delta}}^2 \xbar{\Delta}^\dagger \xbar{\Delta} + B_\Delta \left( \Delta \xbar{\Delta} + \text{h.c.}\right).
\end{equation}
For $m_\Delta^2 = m_{\bar{\Delta}}^2$ and $m_S^2 > 0$, there is a potential minimum with $v_\Delta = v_{\bar{\Delta}}$, $\left<S\right> = 0$, which satisfies the minimization condition
\begin{equation}
\frac{1}{2}\lambda^2 v_\Delta^2  = \lambda f^2 - m_\Delta^2 - B_\Delta.\label{equ:minimization}
\end{equation}
Integrating out the heavy field now results in
\begin{equation}
V_{D, \text{eff}} \supset  \frac{1}{8}\left(g^2 + \xi g_R^2 \right) \left(\left|H_u^0\right|^2 -  \left|H_d^0\right|^2\right)^2, ~~~~~~ \xi =1-\frac{g_R^2}{g_R^2 + g_X^2 + \frac{m_\Delta^2}{v_\Delta^2}}.
\end{equation}
This interpolates between the decoupling and non-decoupling limits, $g'^2 \leq \xi g_R^2 \leq g_R^2$. We see that the non-decoupling limit, $m_\Delta^2/v_\Delta^2 \to \infty$, can only be achieved at the expense of a fine-tuned cancellation between terms on the RHS of \Equ{equ:minimization}. A crude fine tuning measure can be defined by $\Delta_\text{FT} \equiv 2 m_\Delta^2 / (\lambda^2 v_\Delta^2)$. For $\lambda^2 \sim 1$, $m_\Delta^2/v_\Delta^2 \simeq 2.5$ is compatible with $\Delta_\text{FT} \sim 5$. There will also be a fine tuning associated with the sensitivity of the EWSB Higgs soft masses to $m_\Delta$, but this arises only at two loops \cite{Martin:1993zk}
\begin{equation}
\frac{d m_\Phi^2}{d \log \mu} \sim \frac{g_R^4}{16 \pi^4} m_\Delta^2 ~~~~~~ \left(\xbar{DR}\right).
\end{equation}
This contribution to the fine tuning of the EWSB Higgs potential is usually subdominant compared to that associated with the RH gauge symmetry scale, and so we will neglect it here.

\section{Flavour constraints: Additional details}

In this appendix, we provide more details of the convention used in our flavour analysis, and also present the constraints from other FCNC contributions that we have analysed. Note that these constraints are significantly weaker than that from tree-level $Z'$ presented in the main text.

\subsection{Down-type quark masses and mixing}

Here, we introduce the conventions we have adopted for down-type quark masses and mixing. The full $6 \times 6$ down-type quark mass matrix is given by $\mathcal{M}$, where
\begin{equation}
\begin{aligned}
\mathcal{L} &\supset -\begin{pmatrix} d_R^{\prime c} & D^{\prime c} \end{pmatrix} \mathcal{M} \begin{pmatrix} d'_L \\ D' \end{pmatrix} + \text{h.c.},\\
\mathcal{M} &= \begin{pmatrix} \frac{v_d}{\sqrt{2}}y' & \frac{v_D}{\sqrt{2}}z \\ 0 & M \end{pmatrix}.
\end{aligned}
\end{equation}
We have introduced a new Yukawa matrix $y'$ which in general differs from $y$. This is motivated by the need to modify the tree-level mass matrix as suggested in \Sec{sec:exotic} to obtain the correct light quark mass relations. The origin of such a modification will be discussed later.

We first perform block-diagonalisation of the mass matrix before EWSB, i.e. with $v_d = 0$. No transformation of the LH quarks is required, while the RH quarks transform as
\begin{equation}
\begin{pmatrix} d^{\prime c}_R \\ D^{\prime c} \end{pmatrix}
\equiv \begin{pmatrix} c_R & -\tilde{s}_R \\ s_R & \tilde{c}_R \end{pmatrix}^* \begin{pmatrix} d^{\prime\prime c}_R \\ D^{\prime\prime c} \end{pmatrix},
\end{equation}
where $d^{\prime\prime c}_R$ and $D^{\prime\prime c}$ are intermediate basis. In this basis, the full mass matrix becomes
\begin{equation}
\begin{aligned}
\mathcal{M}' &= \begin{pmatrix} 0 & 0 \\ 0 & M_D \end{pmatrix},\\
M_D &\equiv -\frac{v_D}{\sqrt{2}} \tilde{s}_R^\dagger z + \tilde{c}_R^\dagger M.
\end{aligned}
\end{equation}

We reintroduce the EWSB masses, so $\mathcal{M}'$ is no longer block-diagonal
\begin{equation}
\mathcal{M}' = \begin{pmatrix} \frac{v_d}{\sqrt{2}}c_R^\dagger y' & 0 \\ -\frac{v_d}{\sqrt{2}}\tilde{s}_R^\dagger y' & M_D \end{pmatrix}.
\end{equation}
Due to the hierarchy between the EWSB masses and $M_D$, we can use the see-saw formula for block-diagonalisation. We define $\epsilon \equiv |v_d|/m_D$, where $m_D$ is the characteristic eigenvalue size of $M_D$. The LH quarks now transform as
\begin{equation}
\begin{aligned}
\begin{pmatrix} d'_L \\ D' \end{pmatrix}
&\equiv \begin{pmatrix} c_L & -s_L^{\dagger} \\ s_L & \tilde{c}_L \end{pmatrix} \begin{pmatrix} d''_L \\ D'' \end{pmatrix}\\
&\approx
\begin{pmatrix} 1 - \frac{|v_d|^2}{4} y^{\prime\dagger} \tilde{s}_R (M_D M_D^\dagger)^{-1} \tilde{s}_R^\dagger y' &
-\frac{v_d^*}{\sqrt{2}}y^{\prime\dagger} \tilde{s}_R (M_D^\dagger)^{-1} \\
\frac{v_d}{\sqrt{2}} M_D^{-1} \tilde{s}_R^\dagger y' &
1 - \frac{|v_d|^2}{4} M_D^{-1}\tilde{s}_R^\dagger y' y^{\prime\dagger} \tilde{s}_R (M_D^\dagger)^{-1} \end{pmatrix}
\begin{pmatrix} d''_L \\ D'' \end{pmatrix},
\end{aligned}
\end{equation}
with mixing angles of order $\mathcal{O}(\epsilon)$. The RH quarks also transform but with much smaller mixing angles of order $\mathcal{O}(\epsilon^2)$, which we ignore for now. The full $6\times6$ mass matrix becomes
\begin{equation}
\begin{aligned}
\mathcal{M}'' &\approx \begin{pmatrix} M_d & 0 \\ 0 & M_D \end{pmatrix},\\
M_d &\equiv \frac{v_d}{\sqrt{2}} c_R^\dagger y'.
\end{aligned}
\end{equation}
$M_d$ can be thought of as the $3\times3$ mass matrix for $d''_L$ and $d^{\prime\prime c}_R$, and $M_D$ for $D''$ and $D^{\prime \prime c}$. We now perform $3\times 3$ unitary transformations $U_L^d$, $(U_R^d)^*$, $U_L^D$ and $(U_R^D)^*$ on the intermediate basis to diagonalise these mass matrices. Combining all the transformations, we find the following relation between the interaction basis and the mass basis:
\begin{equation}
\begin{aligned}
\begin{pmatrix} d'_L \\ D' \end{pmatrix} &= \begin{pmatrix} c_L & -s_L^{\dagger} \\ s_L & \tilde{c}_L \end{pmatrix} \begin{pmatrix} U_L^d d_L \\ U_L^D D \end{pmatrix},\\
\begin{pmatrix} d^{\prime c}_R \\ D^{\prime c} \end{pmatrix} &= \begin{pmatrix} c_R & -\tilde{s}_R \\ s_R & \tilde{c}_R \end{pmatrix}^* \begin{pmatrix} (U_R^d)^* d_R^c \\ (U_R^D)^* D^c \end{pmatrix}.\\
\end{aligned}
\end{equation}
For example, $c_L U_L^d$ can be identified with the usual CKM matrix $V^L_{\text{CKM}}$, and $c_R U_R^d$ with the RH analogue $V^R_{\text{CKM}}$.

We now discuss the quark mass spectrum. Generic LRS models require that the quarks couple through two sets of Yukawa couplings to the bidoublet Higgs $\Phi$ and its conjugate $\tilde{\Phi}$, to generate the correct up- and down-type mass spectrum. In our model however, the coupling to $\tilde{\Phi}$ is forbidden by the holomorphy of the superpotential, so we only have a single set of couplings $y$. In the up-type mass basis, we expect that $y = \sqrt{2} M_u^{\text{diag}}/v_u$, where $M_u^{\text{diag}}$ is the diagonalised up-type mass matrix. Meanwhile, due to the mixing between $d_R^c$ and $D^c$, the down-type mass matrix becomes $c_R^\dagger y v_d/\sqrt{2}$, so a suitable choice of the matrix $c_R$ should in principle reproduce the correct down-type mass matrix. For example, one can reproduce the correct strange and bottom masses $m_s$ and $m_b$ given $c_R$ of the form
\begin{equation}
c_R \approx \begin{pmatrix} c_d & 0 & 0 \\ 0 & c_d & 0 \\ 0 & 0 & c_b \end{pmatrix},
\end{equation}
with the appropriate values of $c_d$ and $\tan\beta$ taken from, say, \Fig{fig:maxhiggsmass}. We have chosen the first two diagonal elements of $c_R$ to be the same to avoid flavour issues, which we elaborate later. However, the down quark mass $m_d$ always ends up too small, even if we now increase the first diagonal element from $c_d$ to 1. As mentioned in \Sec{sec:exotic}, one solution is to introduce nonrenormalisable operators that can contribute to the down-type mass matrix, analogous to the approach used in \cite{Guadagnoli:2010sd} for up-type quarks. This is equivalent to adding to $y$ a generic matrix of size $\mathcal{O}\left(\sqrt{2} m_u'/v_u \right)$, where we have defined $m_u' \equiv m_c m_d/m_s$. The modified matrix, which we denote as $y'$, remains approximately diagonal and hierarchical:
\begin{equation}
y' \approx \frac{\sqrt{2}}{v_u} \begin{pmatrix} \mathcal{O}(m_u') & \mathcal{O}(m_u') & \mathcal{O}(m_u')\\
\mathcal{O}(m_u') & m_c & \mathcal{O}(m_u')\\
\mathcal{O}(m_u') & \mathcal{O}(m_u') & m_t
\end{pmatrix}.
\label{eq:form-of-yukawa}
\end{equation}
We leave the feasibility study of such a modification to future work. We note that it may also be possible to obtain the correct quark mass spectrum through loop effects involving the SUSY-breaking terms \cite{Babu:1998tm}.

There are various attractive features associated with having $y'$ of the form given in \Equ{eq:form-of-yukawa}. First, as we shall see later, it helps to alleviate some of the FCNC constraints on the model. Second, since $U_L^d$ is the transformation that diagonalises $y^{\prime\dagger} c_R c_R^\dagger y'$, and since $c_L$ deviates from identity only by $\mathcal{O}(\epsilon^2)$, the form of $y'$ also ensures that $U_L^d$ and hence $V_{\text{CKM}}^L$ is close to identity with only small mixing angles, in agreement with measurements. Finally, we note that the strongest constraint on CKM unitarity comes from the experimental measurements \cite{GonzalezAlonso:2011tp}
\begin{equation}
\Delta_{\text{CKM}} \equiv |(V_{\text{CKM}}^L)_{ud}|^2 + |(V_{\text{CKM}}^L)_{us}|^2 + |(V_{\text{CKM}}^L)_{ub}|^2 - 1 = (-1 \pm 6) \times 10^{-4}.
\end{equation}
In the model, $\Delta_{\text{CKM}}$ is suppressed both by a factor of $\mathcal{O}(\epsilon^2)$ as well as the small elements of $y'$ and so satisfy the unitarity constraints.

\subsection{Tree-level FCNCs}

\subsubsection{Higgses}

In generic LRS models, due to the quarks coupling to both $\Phi$ and $\tilde{\Phi}$, one linear combination of the neutral Higgs can generate tree-level FCNCs, which in turn constrains its mass to more than $10 \mbox{ TeV}$. In supersymmetric model, the coupling to $\tilde{\Phi}$ is forbidden due to holomorphy; however, the issue of tree-level Higgs FCNC still lingers in the down-type sector due to the mixing with vector-like quarks. In particular, we consider the quark coupling $\mathcal{C}^{\text{tree}}_{hd}$ to the neutral down-type Higgs
\begin{equation}
\begin{aligned}
\mathcal{L} &\supset -\frac{1}{\sqrt{2}}d^{\prime c}_R y' d'_L h_d^0 + \text{h.c} \\
&= - d_R^c  \mathcal{C}^{\text{tree}}_{h_d} d_L h_d^0 + (\text{terms involving $D$ and $D^c$}) + \text{h.c.},\\
\mathcal{C}^{\text{tree}}_{h_d} &\equiv V_{\text{CKM}}^{R\dagger} y' V_{\text{CKM}}^L\\
&= \frac{1}{v_d} M_d^{\text{diag}} - \frac{v_d^*}{4} M_d^{\text{diag}} U_L^{d\dagger} y^{\prime\dagger} \tilde{s}_R (M_D M_D^\dagger)^{-1} \tilde{s}_R^\dagger y' U_L^d,
\end{aligned}
\end{equation}
where $M_d^{\text{diag}}$ is the $3\times 3$ diagonal matrix of down-type quark masses. Besides the overall mass suppression of order $\mathcal{O}(\epsilon^2)$, the off-diagonal terms of $\mathcal{C}^{\text{tree}}_{h_d}$ are further suppressed by the fact that $M_d^{\text{diag}}$, $y'$ and $U_L^d$ are diagonal and/or hierarchical. As a result, the $|\Delta F|=2$ FCNC contributions from this coupling turns out to be negligible.

Another source of tree-level FCNC is the down-type quark coupling to the neutral component of the RH Higgs doublet $H_R$
\begin{equation}
\begin{aligned}
\mathcal{L} &\supset -\frac{1}{\sqrt{2}}d^{\prime c}_R z d'_L h_R^0 + \text{h.c} \\
&= -d_R^c  \mathcal{C}^{\text{tree}}_{h_R} d_L h_R^0 + (\text{terms involving $D$ and $D^c$}) + \text{h.c.},\\
\mathcal{C}^{\text{tree}}_{h_R} &\equiv V_{\text{CKM}}^{R\dagger} z s_L U_L^d\\
&= v_d V_{\text{CKM}}^{R\dagger} z (M_D)^{-1} \tilde{s}_R^\dagger y' U_L^d.
\end{aligned}
\end{equation}
There is again a mass suppression of order $\mathcal{O}(\epsilon)$, while $y'$ and $U_L^d$ further suppresses off-diagonal couplings except for $d_R^c b_L$ and $s_R^c b_L$. Therefore, the strongest constraints comes from $B_d -\bar{B}_d$ mixing. Assuming experimental bounds on the operator $(d_R^c b_L)^2$ to be comparable to that of $(d_R^c b_L)(\overline{d_L} \overline{b_R^c})$, we find that \cite{Isidori:2010kg}
\begin{equation}
\left(\frac{3 \mbox{ TeV}}{m_{H_R}}\right)^2 \left(\frac{2 \mbox{ TeV}}{m_D}\right)^2 \left(\frac{4}{\tan\beta}\right)^2
\left( \frac{(V_{\text{CKM}}^{R\dagger}z(\tilde{M}_D)^{-1}\tilde{s}_R^\dagger)_{13}}{0.4}\right)^2 \lesssim 0.03.
\end{equation}
where we have defined $\tilde{M}_D = M_D /m_D$ so that it is a generic $\mathcal{O}(1)$ matrix. The reference value of 
0.4 for $(V_{\text{CKM}}^{R\dagger}z(\tilde{M}_D)^{-1}\tilde{s}_R^\dagger)_{13}$ assumes $c_R$ to be a generic matrix of order $\mathcal{O}(c_d)$, and all other matrices of order $\mathcal{O}(1)$.

The constraint above seems to imply the need for some suppression of the relevant 13 element. However, one finds from numerical simulations with generic $z$ and $M$ that this element is almost always already smaller than what is required above. A brief explanation goes as follows: First, since $U_R^d$ and $U_L^d$ diagonalises $M_D$, we have $V_{\text{CKM}}^{R\dagger} y' U_L^d = M_d^{\text{diag}}$, which implies that the 12 and 13 elements of $V_{\text{CKM}}^{R\dagger}$ are necessarily small. Second, the $\mathcal{O}(c_d)$ hierarchy between $\frac{v_D}{\sqrt{2}}z$ and $M$ results in the combination $z(\tilde{M}_D)^{-1}\tilde{s}_R^\dagger$ being roughly diagonal. Combining both effects, we find the relevant 13 element to be much smaller than the generic size.

\subsubsection{Neutral gauge bosons}

We now consider tree-level FCNCs from $Z$ and $Z'$. We work in the basis before $Z-Z'$ mixing and regard the mixing as a perturbative mass insertion, in which case the couplings to $Z$ and $Z'$ are simply $(g/c_w)(T_L^3 - Q_{\text{EM}} s_w^2)$ and $(g s_w/c_w)(T_R^3/t_{w'} - Q_X t_{w'})$ respectively, where $t_{w'} \equiv g_X/g_R$.

Since $D'$ and $d'_L$ have different $Z$-couplings, the $Z$-coupling to the mass basis $d_L$ is non-universal and given by the matrix $\mathcal{C}^{\text{tree}}_{Z-d_L}$, defined as
\begin{equation}
\begin{aligned}
\mathcal{L} &\supset  Z^\mu \overline{d_L} \mathcal{C}^{\text{tree}}_{Z-d_L} \bar{\sigma}_\mu d_L,\\
\mathcal{C}^{\text{tree}}_{Z-d_L} &\equiv \frac{g}{c_w} \left[ \frac{1}{3}s_w^2-\frac{1}{2} V_{\text{CKM}}^{L\dagger} V_{\text{CKM}}^L \right]\\
&= \frac{g}{c_w} \left[ \frac{1}{3}s_w^2-\frac{1}{2} + \frac{|v_d|^2}{4} U_L^{d\dagger} y^{\prime\dagger} \tilde{s}_R (M_D M_D^\dagger)^{-1} \tilde{s}_R^\dagger y' U_L^d \right].
\end{aligned}
\end{equation}
Besides the mass suppression of order $\mathcal{O}(\epsilon^2)$, the off-diagonal terms in $\mathcal{C}^{\text{tree}}_{Z-d_L}$ is further suppressed by $y'$ and $U_L^d$. As a result, their contributions to $|\Delta F|=2$ processes turns out to be negligible. A similar argument can be made for $Z'$-couplings to $d_L$.

We now move on to $d_R^c$. Since both $d^{\prime c}_R$ and $D^{\prime c}$ have the same couplings to $Z$, there is no tree-level FCNC mediated by $Z$. The FCNC mediated by $Z'$ has already been discussed in the main text.

\subsection{One-loop FCNCs}

\begin{figure}[t]
\begin{center} 
\begin{tikzpicture} 
\draw[fwfermion] (0,1) -- (4,1) node[pos=0,left] {$c$} node[right] {$u$} node[midway,above] {$D_i$}; 
\draw[bwfermion] (0,-1) -- (4,-1) node[pos=0,left] {$\bar{u}$} node[right] {$\bar{c}$} node[midway,below] {$\bar{D}_j$}; 
\draw[vector] (1,-1) -- (1,1) node[midway,left] {$W_{L/R}$}; 
\draw[vector] (3,-1) -- (3,1) node[midway,right] {$W_{L/R}$}; 
\end{tikzpicture} 
\quad
\begin{tikzpicture} 
\draw[fwfermion] (0,1) -- (4,1) node[pos=0,left] {$s$} node[right] {$d$} node[midway,above] {$D_i$}; 
\draw[bwfermion] (0,-1) -- (4,-1) node[pos=0,left] {$\bar{d}$} node[right] {$\bar{s}$} node[midway,below] {$\bar{D}_j$}; 
\draw[scalar] (1,-1) -- (1,1) node[midway,left] {$h_R^0$}; 
\draw[scalar] (3,-1) -- (3,1) node[midway,right] {$h_R^0$}; 
\end{tikzpicture}
\end{center}
\caption{Examples of $|\Delta F| = 2$ box diagrams for $D-\bar{D}$ and $K-\bar{K}$ mixings.}
\label{fig:box-diag}
\end{figure}
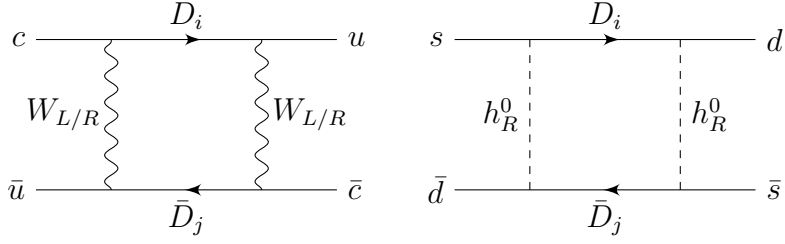

Numerous box diagrams in our model can contribute to $|\Delta F| = 2$ processes. Besides those from LRS and vector-like quarks, we also have additional diagrams involving the superpartners. A complete analysis of all such box diagrams and interference lies beyond the scope of this work, and we will only consider a small subset of diagrams involving the new quarks as shown in \Fig{fig:box-diag}.

\subsubsection{$W_{L/R}-W_{L/R}$ box diagrams for $D-\bar{D}$ mixing}

The $W_{L/R}$ gauge couplings of interest are given by
\begin{equation}
\begin{aligned}
\mathcal{L} &\supset \frac{g}{\sqrt{2}} W_L^{+\mu} \overline{u_L} (-s_L^\dagger U_L^D) \bar{\sigma}_\mu D - \frac{g_R}{\sqrt{2}} W_R^{+\mu} \overline{D_R^c} (-\tilde{s}_R U_R^D)^T \bar{\sigma}_\mu u_R^c + \text{h.c}.
\end{aligned}
\end{equation}
Due to the factors of $\mathcal{O}(\epsilon)$ and $y'$ present in $s_L$, it turns out that the $W_L-W_L$ and $W_L-W_R$ contributions are highly suppressed, so only the $W_R-W_R$ contributions are of interest. The box diagram can in principle be evaluated using the Inami-Lim formula \cite{Inami:1980fz}. However, we will make a pessimistic approximation, from which we obtain the effective Hamiltonian
\begin{equation}
\mathcal{H}^{\text{eff}}_{W_R}(|\Delta C| = 2) \approx \frac{g_R^4}{128\pi^2} \frac{m_D^2}{m_{W'}^4} \left(\max_{i}[ (\tilde{s}_R U_R^D)_{ui}(\tilde{s}_R U_R^D)_{ci}^*]\right)^2(\overline{c_R^c} \bar{\sigma}_\mu u_R^c)^2 + \text{h.c}.
\end{equation}
To satisfy the bounds on $\Delta m_D$, we require that
\begin{equation}
\left(\frac{g_R/g}{1.2}\right)^4 \left( \frac{m_D}{2 \mbox{ TeV}} \right)^2 \left(\max_{i}[ (\tilde{s}_R U_R^D)_{ui}(\tilde{s}_R U_R^D)_{ci}^*]\right)^2 \lesssim 0.01.
\end{equation}
We see that we only require a small suppression of the off-diagonal $\tilde{s}_R U_R^D$ elements since they appear here to the fourth power.

\subsubsection{Box diagrams involving $H_R$}

We have chosen to consider box diagrams involving $H_R$ instead of those involving $\Phi$, since we expect constraints from the latter to be weaker due to $y'$ being hierarchical and nearly diagonal. The relevant couplings are given by
\begin{equation}
\mathcal{L}\supset -\frac{1}{\sqrt{2}} d_R^c (V_{\text{CKM}}^{R\dagger} z \tilde{c}_L V_L) D h_R^0 + \text{h.c.}
\end{equation}
The formula for the loop integral can be obtained from \cite{Barger:1989fj}, although we will again make a pessimistic approximations. We then obtain the effective Hamiltonian
\begin{equation}
\mathcal{H}^{\text{eff}}_{H_R}(|\Delta S| = 2) \approx \frac{1}{512\pi^2} \frac{1}{m_{h_R}^2} \left(\max_{i}[ (V_{\text{CKM}}^R z\tilde{c}_L U_L^D)_{si}(V_{\text{CKM}}^R z\tilde{c}_L U_L^D)_{di}^*]\right)^2(\overline{d_R^c} \bar{\sigma}_\mu s_R^c)^2 + \text{h.c.}
\end{equation}
for $K-\bar{K}$ mixing, from which we require that \cite{Isidori:2010kg}
\begin{equation}
\left(\frac{3 \mbox{ TeV}}{m_{h_R}^2}\right)^2 \left(\frac{\max_{i}[ (V_{\text{CKM}}^R z\tilde{c}_L U_L^D)_{si}(V_{\text{CKM}}^R z\tilde{c}_L U_L^D)_{di}^*]}{0.2}\right)^2 \lesssim 1.
\end{equation}
The reference value of 0.2 is again based on regarding $c_R$ as a generic matrix of order $\mathcal{O}(c_d)$, and all other matrices of order $\mathcal{O}(1)$. We see that the constraint is satisfied without any suppression of the off-diagonal terms. The same holds for box diagrams for $B_{d(s)}-\bar{B}_{d(s)}$ mixing.

\bibliographystyle{JHEP}
\bibliography{Dibosonbib}

\providecommand{\href}[2]{#2}\begingroup\raggedright\begin{thebibliography}{10}

\bibitem{Ellis:1988er}
J.~R. Ellis, J.~F. Gunion, H.~E. Haber, L.~Roszkowski, and F.~Zwirner, {\it
  {Higgs Bosons in a Nonminimal Supersymmetric Model}},  {\em Phys. Rev.} {\bf
  D39} (1989) 844.

\bibitem{Drees:1988fc}
M.~Drees, {\it {Supersymmetric Models with Extended Higgs Sector}},  {\em Int.
  J. Mod. Phys.} {\bf A4} (1989) 3635.

\bibitem{Batra:2003nj}
P.~Batra, A.~Delgado, D.~E. Kaplan, and T.~M.~P. Tait, {\it {The Higgs mass
  bound in gauge extensions of the minimal supersymmetric standard model}},
  {\em JHEP} {\bf 02} (2004) 043,
  [\href{http://arxiv.org/abs/hep-ph/0309149}{{\tt hep-ph/0309149}}].

\bibitem{Maloney:2004rc}
A.~Maloney, A.~Pierce, and J.~G. Wacker, {\it {D-terms, unification, and the
  Higgs mass}},  {\em JHEP} {\bf 06} (2006) 034,
  [\href{http://arxiv.org/abs/hep-ph/0409127}{{\tt hep-ph/0409127}}].

\bibitem{Bertuzzo:2014sma}
E.~Bertuzzo and C.~Frugiuele, {\it {A natural SM-like 126 GeV Higgs via
  non-decoupling D-terms}},  \href{http://arxiv.org/abs/1412.2765}{{\tt
  arXiv:1412.2765}}.

\bibitem{Aad:2015owa}
{\bf ATLAS} Collaboration, G.~Aad et~al., {\it {Search for high-mass diboson
  resonances with boson-tagged jets in proton-proton collisions at $\sqrt{s} =
  8$ TeV with the ATLAS detector}},
  \href{http://arxiv.org/abs/1506.00962}{{\tt arXiv:1506.00962}}.

\bibitem{Allanach:2015hba}
B.~C. Allanach, B.~Gripaios, and D.~Sutherland, {\it {Anatomy of the ATLAS
  diboson anomaly}},  {\em Phys. Rev.} {\bf D92} (2015), no.~5 055003,
  [\href{http://arxiv.org/abs/1507.01638}{{\tt arXiv:1507.01638}}].

\bibitem{Fichet:2015yia}
S.~Fichet and G.~von Gersdorff, {\it {Effective theory for neutral resonances
  and a statistical dissection of the ATLAS diboson excess}},
  \href{http://arxiv.org/abs/1508.04814}{{\tt arXiv:1508.04814}}.

\bibitem{Aad:2015ufa}
{\bf ATLAS} Collaboration, G.~Aad et~al., {\it {Search for production of
  $WW/WZ$ resonances decaying to a lepton, neutrino and jets in $pp$ collisions
  at $\sqrt{s}=8$ TeV with the ATLAS detector}},  {\em Eur. Phys. J.} {\bf C75}
  (2015), no.~5 209, [\href{http://arxiv.org/abs/1503.04677}{{\tt
  arXiv:1503.04677}}]. [Erratum: Eur. Phys. J.C75,370(2015)].

\bibitem{Aad:2014xka}
{\bf ATLAS} Collaboration, G.~Aad et~al., {\it {Search for resonant diboson
  production in the $\mathrm {\ell \ell }q\bar{q}$ final state in $pp$
  collisions at $\sqrt{s} = 8$ TeV with the ATLAS detector}},  {\em Eur. Phys.
  J.} {\bf C75} (2015) 69, [\href{http://arxiv.org/abs/1409.6190}{{\tt
  arXiv:1409.6190}}].

\bibitem{ATLASdibosoncomb}
T.~A. collaboration, {\it {Search for $WW$, $WZ$, and $ZZ$ resonances in $pp$
  collisions at $\sqrt{s} = 8$ \tev with the ATLAS detector}}, .

\bibitem{Khachatryan:2014hpa}
{\bf CMS} Collaboration, V.~Khachatryan et~al., {\it {Search for massive
  resonances in dijet systems containing jets tagged as W or Z boson decays in
  pp collisions at $ \sqrt{s} $ = 8 TeV}},  {\em JHEP} {\bf 08} (2014) 173,
  [\href{http://arxiv.org/abs/1405.1994}{{\tt arXiv:1405.1994}}].

\bibitem{Khachatryan:2014gha}
{\bf CMS} Collaboration, V.~Khachatryan et~al., {\it {Search for massive
  resonances decaying into pairs of boosted bosons in semi-leptonic final
  states at $\sqrt{s} =$ 8 TeV}},  {\em JHEP} {\bf 08} (2014) 174,
  [\href{http://arxiv.org/abs/1405.3447}{{\tt arXiv:1405.3447}}].

\bibitem{Khachatryan:2015sja}
{\bf CMS} Collaboration, V.~Khachatryan et~al., {\it {Search for resonances and
  quantum black holes using dijet mass spectra in proton-proton collisions at
  $\sqrt{s} =$ 8 TeV}},  {\em Phys. Rev.} {\bf D91} (2015), no.~5 052009,
  [\href{http://arxiv.org/abs/1501.04198}{{\tt arXiv:1501.04198}}].

\bibitem{Aad:2014aqa}
{\bf ATLAS} Collaboration, G.~Aad et~al., {\it {Search for new phenomena in the
  dijet mass distribution using $p-p$ collision data at $\sqrt{s}=8$ TeV with
  the ATLAS detector}},  {\em Phys. Rev.} {\bf D91} (2015), no.~5 052007,
  [\href{http://arxiv.org/abs/1407.1376}{{\tt arXiv:1407.1376}}].

\bibitem{Khachatryan:2014dka}
{\bf CMS} Collaboration, V.~Khachatryan et~al., {\it {Search for heavy
  neutrinos and $\mathrm {W}$ bosons with right-handed couplings in
  proton-proton collisions at $\sqrt{s} = 8\,\text {TeV} $}},  {\em Eur. Phys.
  J.} {\bf C74} (2014), no.~11 3149,
  [\href{http://arxiv.org/abs/1407.3683}{{\tt arXiv:1407.3683}}].

\bibitem{Brehmer:2015cia}
J.~Brehmer, J.~Hewett, J.~Kopp, T.~Rizzo, and J.~Tattersall, {\it {Symmetry
  Restored in Dibosons at the LHC?}},
  \href{http://arxiv.org/abs/1507.00013}{{\tt arXiv:1507.00013}}.

\bibitem{Hisano:2015gna}
J.~Hisano, N.~Nagata, and Y.~Omura, {\it {Interpretations of the ATLAS Diboson
  Resonances}},  {\em Phys. Rev.} {\bf D92} (2015), no.~5 055001,
  [\href{http://arxiv.org/abs/1506.03931}{{\tt arXiv:1506.03931}}].

\bibitem{Cheung:2015nha}
K.~Cheung, W.-Y. Keung, P.-Y. Tseng, and T.-C. Yuan, {\it {Interpretations of
  the ATLAS Diboson Anomaly}},  \href{http://arxiv.org/abs/1506.06064}{{\tt
  arXiv:1506.06064}}.

\bibitem{Dobrescu:2015qna}
B.~A. Dobrescu and Z.~Liu, {\it {A W' boson near 2 TeV: predictions for Run 2
  of the LHC}},  \href{http://arxiv.org/abs/1506.06736}{{\tt
  arXiv:1506.06736}}.

\bibitem{Gao:2015irw}
Y.~Gao, T.~Ghosh, K.~Sinha, and J.-H. Yu, {\it {SU(2)×SU(2)×U(1)
  interpretations of the diboson and Wh excesses}},  {\em Phys. Rev.} {\bf D92}
  (2015), no.~5 055030, [\href{http://arxiv.org/abs/1506.07511}{{\tt
  arXiv:1506.07511}}].

\bibitem{Cao:2015lia}
Q.-H. Cao, B.~Yan, and D.-M. Zhang, {\it {Simple Non-Abelian Extensions and
  Diboson Excesses at the LHC}},  \href{http://arxiv.org/abs/1507.00268}{{\tt
  arXiv:1507.00268}}.

\bibitem{Dobrescu:2015yba}
B.~A. Dobrescu and Z.~Liu, {\it {Heavy Higgs bosons and the 2 TeV $W'$ boson}},
   \href{http://arxiv.org/abs/1507.01923}{{\tt arXiv:1507.01923}}.

\bibitem{Krauss:2015nba}
M.~E. Krauss and W.~Porod, {\it {Is the CMS eejj excess a hint for light
  supersymmetry?}},  {\em Phys. Rev.} {\bf D92} (2015), no.~5 055019,
  [\href{http://arxiv.org/abs/1507.04349}{{\tt arXiv:1507.04349}}].

\bibitem{Dev:2015pga}
P.~S.~B. Dev and R.~N. Mohapatra, {\it {Unified explanation of the $eejj$,
  diboson and dijet resonances at the LHC}},
  \href{http://arxiv.org/abs/1508.02277}{{\tt arXiv:1508.02277}}.

\bibitem{Coloma:2015una}
P.~Coloma, B.~A. Dobrescu, and J.~Lopez-Pavon, {\it {Right-Handed Neutrinos and
  the 2 TeV $W'$ Boson}},  \href{http://arxiv.org/abs/1508.04129}{{\tt
  arXiv:1508.04129}}.

\bibitem{Gluza:2015goa}
J.~Gluza and T.~Jeliński, {\it {Heavy neutrinos and the pp→lljj CMS data}},
  {\em Phys. Lett.} {\bf B748} (2015) 125--131,
  [\href{http://arxiv.org/abs/1504.05568}{{\tt arXiv:1504.05568}}].

\bibitem{Deppisch:2015cua}
F.~F. Deppisch, L.~Graf, S.~Kulkarni, S.~Patra, W.~Rodejohann, N.~Sahu, and
  U.~Sarkar, {\it {Reconciling the 2 TeV Excesses at the LHC in a Linear Seesaw
  Left-Right Model}},  \href{http://arxiv.org/abs/1508.05940}{{\tt
  arXiv:1508.05940}}.

\bibitem{Pati:1974yy}
J.~C. Pati and A.~Salam, {\it {Lepton Number as the Fourth Color}},  {\em Phys.
  Rev.} {\bf D10} (1974) 275--289. [Erratum: Phys. Rev.D11,703(1975)].

\bibitem{Mohapatra:1974hk}
R.~N. Mohapatra and J.~C. Pati, {\it {Left-Right Gauge Symmetry and an
  Isoconjugate Model of CP Violation}},  {\em Phys. Rev.} {\bf D11} (1975)
  566--571.

\bibitem{Keung:1983uu}
W.-Y. Keung and G.~Senjanovic, {\it {Majorana Neutrinos and the Production of
  the Right-handed Charged Gauge Boson}},  {\em Phys. Rev. Lett.} {\bf 50}
  (1983) 1427.

\bibitem{Zhang:2008jm}
Y.~Zhang, H.~An, X.-d. Ji, and R.~N. Mohapatra, {\it {Light Higgs Mass Bound in
  SUSY Left-Right Models}},  {\em Phys. Rev.} {\bf D78} (2008) 011302,
  [\href{http://arxiv.org/abs/0804.0268}{{\tt arXiv:0804.0268}}].

\bibitem{Babu:2014vba}
K.~S. Babu and A.~Patra, {\it {Higgs Boson Spectra in Supersymmetric Left-Right
  Models}},  \href{http://arxiv.org/abs/1412.8714}{{\tt arXiv:1412.8714}}.

\bibitem{Hewett:1988xc}
J.~L. Hewett and T.~G. Rizzo, {\it {Low-Energy Phenomenology of Superstring
  Inspired E(6) Models}},  {\em Phys. Rept.} {\bf 183} (1989) 193.

\bibitem{Deshpande:1988qq}
N.~G. Deshpande, J.~A. Grifols, and A.~Mendez, {\it {Signatures of Right-handed
  Gauge Bosons Through Their Decays Into $W$ and $Z$ Bosons in High-energy $p
  p$ Collisions}},  {\em Phys. Lett.} {\bf B208} (1988) 141. [Erratum: Phys.
  Lett.B214,661(1988)].

\bibitem{Babu:1998tm}
K.~S. Babu, B.~Dutta, and R.~N. Mohapatra, {\it {Partial Yukawa unification and
  a supersymmetric origin of flavor mixing}},  {\em Phys. Rev.} {\bf D60}
  (1999) 095004, [\href{http://arxiv.org/abs/hep-ph/9812421}{{\tt
  hep-ph/9812421}}].

\bibitem{Guadagnoli:2010sd}
D.~Guadagnoli and R.~N. Mohapatra, {\it {TeV Scale Left Right Symmetry and
  Flavor Changing Neutral Higgs Effects}},  {\em Phys. Lett.} {\bf B694} (2011)
  386--392, [\href{http://arxiv.org/abs/1008.1074}{{\tt arXiv:1008.1074}}].

\bibitem{Aad:2015tba}
{\bf ATLAS} Collaboration, G.~Aad et~al., {\it {Search for pair production of a
  new heavy quark that decays into a $W$ boson and a light quark in $pp$
  collisions at $\sqrt{s} = 8$ TeV with the ATLAS detector}},
  \href{http://arxiv.org/abs/1509.04261}{{\tt arXiv:1509.04261}}.

\bibitem{Aad:2015kqa}
{\bf ATLAS} Collaboration, G.~Aad et~al., {\it {Search for production of
  vector-like quark pairs and of four top quarks in the lepton-plus-jets final
  state in $pp$ collisions at $\sqrt{s}=8$ TeV with the ATLAS detector}},  {\em
  JHEP} {\bf 08} (2015) 105, [\href{http://arxiv.org/abs/1505.04306}{{\tt
  arXiv:1505.04306}}].

\bibitem{Aad:2015gdg}
{\bf ATLAS} Collaboration, G.~Aad et~al., {\it {Analysis of events with
  $b$-jets and a pair of leptons of the same charge in $pp$ collisions at
  $\sqrt{s}=8$ TeV with the ATLAS detector}},
  \href{http://arxiv.org/abs/1504.04605}{{\tt arXiv:1504.04605}}.

\bibitem{Aad:2014efa}
{\bf ATLAS} Collaboration, G.~Aad et~al., {\it {Search for pair and single
  production of new heavy quarks that decay to a $Z$ boson and a
  third-generation quark in $pp$ collisions at $\sqrt{s}=8$ TeV with the ATLAS
  detector}},  {\em JHEP} {\bf 11} (2014) 104,
  [\href{http://arxiv.org/abs/1409.5500}{{\tt arXiv:1409.5500}}].

\bibitem{Aad:2015mba}
{\bf ATLAS} Collaboration, G.~Aad et~al., {\it {Search for vectorlike $B$
  quarks in events with one isolated lepton, missing transverse momentum and
  jets at $\sqrt{s}=$ 8 TeV with the ATLAS detector}},  {\em Phys. Rev.} {\bf
  D91} (2015), no.~11 112011, [\href{http://arxiv.org/abs/1503.05425}{{\tt
  arXiv:1503.05425}}].

\bibitem{Khachatryan:2015gza}
{\bf CMS} Collaboration, V.~Khachatryan et~al., {\it {Search for pair-produced
  vector-like B quarks in proton-proton collisions at $\sqrt{s}$ = 8 TeV}},
  \href{http://arxiv.org/abs/1507.07129}{{\tt arXiv:1507.07129}}.

\bibitem{Ball:2012cx}
R.~D. Ball et~al., {\it {Parton distributions with LHC data}},  {\em Nucl.
  Phys.} {\bf B867} (2013) 244--289,
  [\href{http://arxiv.org/abs/1207.1303}{{\tt arXiv:1207.1303}}].

\bibitem{Aad:2014cka}
{\bf ATLAS} Collaboration, G.~Aad et~al., {\it {Search for high-mass dilepton
  resonances in pp collisions at $\sqrt{s}=8$  TeV with the ATLAS
  detector}},  {\em Phys. Rev.} {\bf D90} (2014), no.~5 052005,
  [\href{http://arxiv.org/abs/1405.4123}{{\tt arXiv:1405.4123}}].

\bibitem{Khachatryan:2014fba}
{\bf CMS} Collaboration, V.~Khachatryan et~al., {\it {Search for physics beyond
  the standard model in dilepton mass spectra in proton-proton collisions at $
  \sqrt{s}=8 $ TeV}},  {\em JHEP} {\bf 04} (2015) 025,
  [\href{http://arxiv.org/abs/1412.6302}{{\tt arXiv:1412.6302}}].

\bibitem{Cacciapaglia:2006pk}
G.~Cacciapaglia, C.~Csaki, G.~Marandella, and A.~Strumia, {\it {The Minimal Set
  of Electroweak Precision Parameters}},  {\em Phys. Rev.} {\bf D74} (2006)
  033011, [\href{http://arxiv.org/abs/hep-ph/0604111}{{\tt hep-ph/0604111}}].

\bibitem{Carpentier:2010ue}
M.~Carpentier and S.~Davidson, {\it {Constraints on two-lepton, two quark
  operators}},  {\em Eur. Phys. J.} {\bf C70} (2010) 1071--1090,
  [\href{http://arxiv.org/abs/1008.0280}{{\tt arXiv:1008.0280}}].

\bibitem{CMS-DP-2015-039}
{\bf CMS Collaboration} Collaboration, {\it {Event Display of a Candidate
  Electron-Positron Pair with an Invariant Mass of 2.9 TeV}}, .

\bibitem{Vega:2015fna}
J.~P. Vega and G.~Villadoro, {\it {SusyHD: Higgs mass Determination in
  Supersymmetry}},  {\em JHEP} {\bf 07} (2015) 159,
  [\href{http://arxiv.org/abs/1504.05200}{{\tt arXiv:1504.05200}}].

\bibitem{Heinemeyer:1998yj}
S.~Heinemeyer, W.~Hollik, and G.~Weiglein, {\it {FeynHiggs: A Program for the
  calculation of the masses of the neutral CP even Higgs bosons in the MSSM}},
  {\em Comput. Phys. Commun.} {\bf 124} (2000) 76--89,
  [\href{http://arxiv.org/abs/hep-ph/9812320}{{\tt hep-ph/9812320}}].

\bibitem{Djouadi:2002ze}
A.~Djouadi, J.-L. Kneur, and G.~Moultaka, {\it {SuSpect: A Fortran code for the
  supersymmetric and Higgs particle spectrum in the MSSM}},  {\em Comput. Phys.
  Commun.} {\bf 176} (2007) 426--455,
  [\href{http://arxiv.org/abs/hep-ph/0211331}{{\tt hep-ph/0211331}}].

\bibitem{Ecker:1985vv}
G.~Ecker and W.~Grimus, {\it {CP Violation and Left-Right Symmetry}},  {\em
  Nucl. Phys.} {\bf B258} (1985) 328--360.

\bibitem{Nishiura:1990hh}
H.~Nishiura, E.~Takasugi, and M.~Tanaka, {\it {Light W(R) and the spontaneous
  CP violation. II}},  {\em Prog. Theor. Phys.} {\bf 85} (1991) 343--354.

\bibitem{Deshpande:1990ip}
N.~G. Deshpande, J.~F. Gunion, B.~Kayser, and F.~I. Olness, {\it {Left-right
  symmetric electroweak models with triplet Higgs}},  {\em Phys. Rev.} {\bf
  D44} (1991) 837--858.

\bibitem{Zhang:2007da}
Y.~Zhang, H.~An, X.~Ji, and R.~N. Mohapatra, {\it {General CP Violation in
  Minimal Left-Right Symmetric Model and Constraints on the Right-Handed
  Scale}},  {\em Nucl. Phys.} {\bf B802} (2008) 247--279,
  [\href{http://arxiv.org/abs/0712.4218}{{\tt arXiv:0712.4218}}].

\bibitem{Maiezza:2010ic}
A.~Maiezza, M.~Nemevsek, F.~Nesti, and G.~Senjanovic, {\it {Left-Right Symmetry
  at LHC}},  {\em Phys. Rev.} {\bf D82} (2010) 055022,
  [\href{http://arxiv.org/abs/1005.5160}{{\tt arXiv:1005.5160}}].

\bibitem{Bjorken:2002vt}
J.~D. Bjorken, S.~Pakvasa, and S.~F. Tuan, {\it {Yet another extension of the
  standard model: Oases in the desert?}},  {\em Phys. Rev.} {\bf D66} (2002)
  053008, [\href{http://arxiv.org/abs/hep-ph/0206116}{{\tt hep-ph/0206116}}].

\bibitem{Golowich:2007ka}
E.~Golowich, J.~Hewett, S.~Pakvasa, and A.~A. Petrov, {\it {Implications of
  $D^0$ - $\bar{D}^0$ Mixing for New Physics}},  {\em Phys. Rev.} {\bf D76}
  (2007) 095009, [\href{http://arxiv.org/abs/0705.3650}{{\tt
  arXiv:0705.3650}}].

\bibitem{Glashow:1976nt}
S.~L. Glashow and S.~Weinberg, {\it {Natural Conservation Laws for Neutral
  Currents}},  {\em Phys. Rev.} {\bf D15} (1977) 1958.

\bibitem{Paschos:1976ay}
E.~A. Paschos, {\it {Diagonal Neutral Currents}},  {\em Phys. Rev.} {\bf D15}
  (1977) 1966.

\bibitem{Zhang:2007qma}
Y.~Zhang, H.~An, and X.-d. Ji, {\it {Constraining the right-handed scale
  through kaon mixing in the supesymmetric left-right model}},  {\em Phys.
  Rev.} {\bf D78} (2008) 035006, [\href{http://arxiv.org/abs/0710.1454}{{\tt
  arXiv:0710.1454}}].

\bibitem{Isidori:2010kg}
G.~Isidori, Y.~Nir, and G.~Perez, {\it {Flavor Physics Constraints for Physics
  Beyond the Standard Model}},  {\em Ann. Rev. Nucl. Part. Sci.} {\bf 60}
  (2010) 355, [\href{http://arxiv.org/abs/1002.0900}{{\tt arXiv:1002.0900}}].

\bibitem{Martin:1993zk}
S.~P. Martin and M.~T. Vaughn, {\it {Two loop renormalization group equations
  for soft supersymmetry breaking couplings}},  {\em Phys. Rev.} {\bf D50}
  (1994) 2282, [\href{http://arxiv.org/abs/hep-ph/9311340}{{\tt
  hep-ph/9311340}}]. [Erratum: Phys. Rev.D78,039903(2008)].

\bibitem{GonzalezAlonso:2011tp}
M.~Gonzalez-Alonso, {\it {New Physics bounds from CKM-unitarity}},  in {\em
  {Proceedings, 45th Rencontres de Moriond on Electroweak Interactions and
  Unified Theories}}, 2011.
\newblock \href{http://arxiv.org/abs/1101.4679}{{\tt arXiv:1101.4679}}.

\bibitem{Inami:1980fz}
T.~Inami and C.~S. Lim, {\it {Effects of Superheavy Quarks and Leptons in
  Low-Energy Weak Processes $K_L \to \mu \bar{\mu}$, $K^+ \to \pi^+ \nu
  \bar{\nu}$ and $K^0 \leftrightarrow \bar{K}^0$}},  {\em Prog. Theor. Phys.}
  {\bf 65} (1981) 297. [Erratum: Prog. Theor. Phys.65,1772(1981)].

\bibitem{Barger:1989fj}
V.~D. Barger, J.~L. Hewett, and R.~J.~N. Phillips, {\it {New Constraints on the
  Charged Higgs Sector in Two Higgs Doublet Models}},  {\em Phys. Rev.} {\bf
  D41} (1990) 3421--3441.

\end{thebibliography}\endgroup

\end{document}